\documentclass[12pt]{article}

\usepackage{graphicx}
\usepackage{amsmath}
\usepackage{a4wide}
\usepackage{bbm}
\usepackage{natbib}
\pdfoutput=1

\newcommand{\ve}{\boldsymbol}
\bibpunct[,]{(}{)}{,}{a}{}{,}

\begin{document}

\title{A unified projection formalism for the Al-Pd-Mn quasicrystal
  $\Xi$-approximants and their metadislocations}

\author{M.~Engel and H.-R.~Trebin\\
  {\normalsize Institut f\"ur Theoretische und Angewandte Physik,}
  {\normalsize Universit\"at Stuttgart,} \\
  {\normalsize Pfaffenwaldring 57, D-70550 Stuttgart, Germany}\\
  {\normalsize email: mengel@itap.physik.uni-stuttgart.de}
}

\maketitle

\begin{abstract}
The approximants $\xi$, $\xi'$ and $\xi'_{n}$ of the quasicrystal Al-Mn-Pd
display most interesting plastic properties as for example phason-induced
deformation processes ({\sc Klein,~H., Audier,~M., Boudard,~M., de
  Boissieu,~M., Beraha,~L.}, and {\sc Duneau,~M.}, 1996, {\it Phil. Mag.}~A,
{\bf 73}, 309.) or metadislocations ({\sc Klein,~H., Feuerbacher,~M.,
  Schall,~P.}, and {\sc Urban,~K.}, 1999, {\it Phys. Rev. Lett.}, {\bf 82},
3468.). Here we demonstrate that the phases and their deformed or defected
states can be described by a simple projection formalism in three-dimensional
space - not as usual in four to six dimensions. With the method we can
interpret microstructures observed with electron microscopy as phasonic phase
boundaries. Furthermore we determine the metadislocations of lowest energy and
relate them uniquely to experimentally observed ones. Since moving
metadislocations in the $\xi'$-phase can create new phason-planes, we suggest
a dislocation induced phase transition from $\xi'$ to $\xi'_{n}$. The methods
developed in this paper can as well be used for various other complex metallic
alloys.
\end{abstract}

\section{Introduction}

A large number of periodic and quasiperiodic phases have been observed in the
ternary Al-Pd-Mn system \cite[]{itapdb:Klein2000a}. Apart from the stable
icosahedral phase (i-phase) there is a stable decagonal phase (d-phase) with
1.2~nm periodicity and a metastable d-phase with 1.6~nm periodicity, found by
\cite{itapdb:Tsai1991a}. The orthorhombic $\Xi$-approximants ($\xi$, $\xi'$
and $\xi'_{n}$) of the i-phase form a class of closely related periodic phases
and can be viewed as approximants of this 1.6~nm d-phase, as they have the
following features in common with the d-phase:
\begin{enumerate}
\item[(a)] They are arrangements of columns of Mackay-type clusters and
  intermediary atoms \cite[]{itapdb:Sun1996a}. These clusters consist of about
  52 atoms, placed on concentric shells of icosahedral symmetry as shown for
  the $\xi'$-phase by \cite{itapdb:Boudard1996b}.
\item[(b)] The columns are in registry, i.e.\ the compounds also have a layer
  structure.
\end{enumerate}

The clusters contain about 80\% of the atoms. Therefore a coarsened structural
description makes sense where only the projections of the cluster columns
along the column lines are marked.  They lead to two-dimensional tilings,
which are characteristic for the respective approximant phase. The vertices of
the tiles correspond to the projections of the cluster columns.

\begin{figure}\begin{center}
\includegraphics[width=16cm]{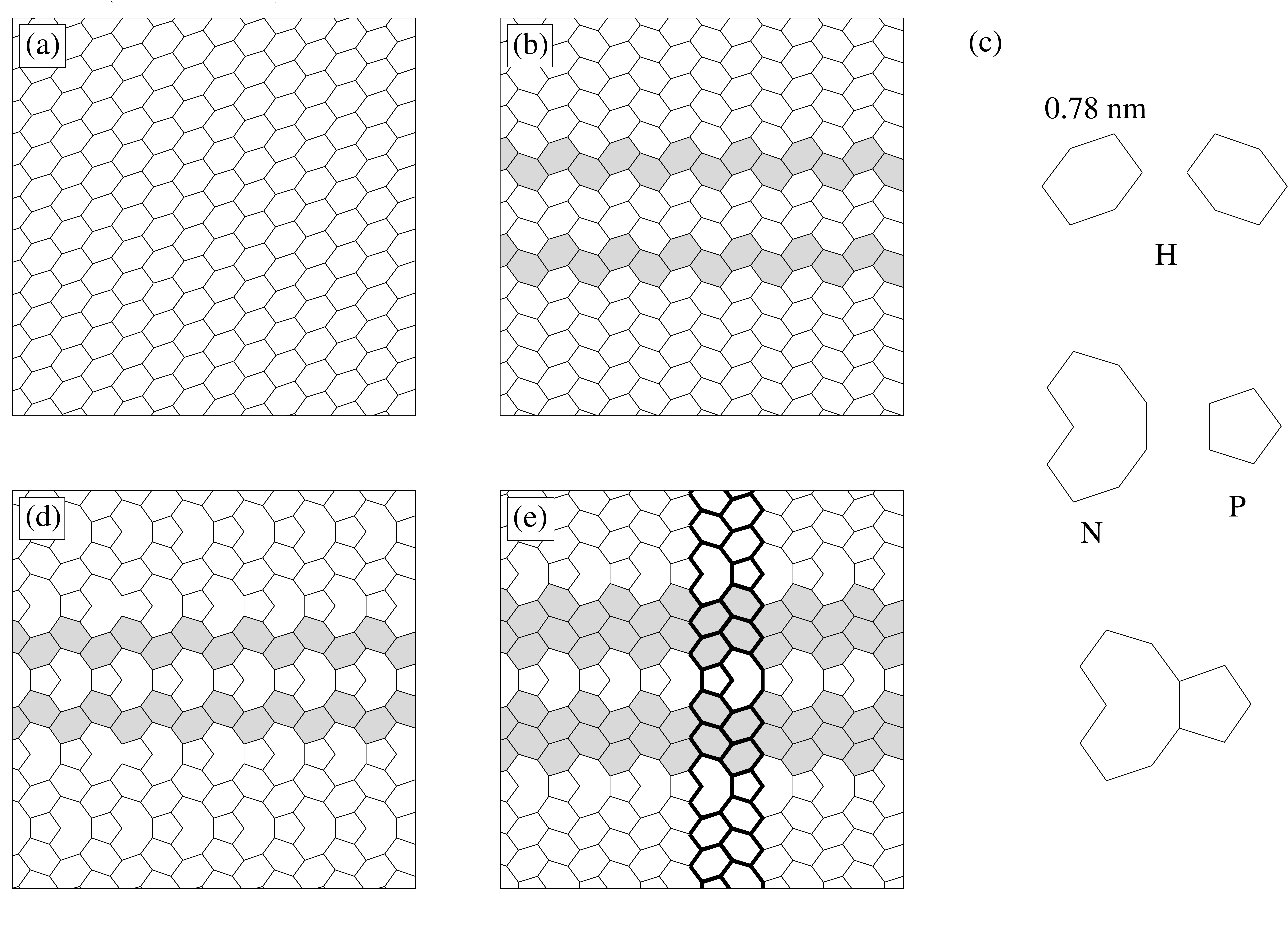}
\caption{Tilings of four different approximant phases of the 1.6~nm
    d-phase. (a)~Parallel alignment of hexagons in the
  $\xi$-phase. (b)~Staggered arrangement of hexagons in the $\xi'$-phase. Two
  hexagon rows are marked grey. (c)~The hexagons H occur in two different
  orientations. The other tiles, the pentagon P and the nonagon N can only be
  observed in combination, called phason-line. The edge length of the tiles
  is: $t^{6D}=0.78$~nm. (d)~In the $\xi_{2}$-phase there is one row of
  hexagons between two phason-lines, also called phason-planes. (e)~In the
  $\xi_{3}$-phase there are two rows of hexagons between neighbouring
  phason-planes. The arrangement of hexagons is flipped at a phason-line as
  highlighted in the figure. \label{fig:tilings}}
\end{center}\end{figure}

The tilings contain flattend hexagons, which for the $\xi$-phase are aligned
parallel (figure~\ref{fig:tilings}(a)), for the $\xi'$-phase are staggered in
two orientations (figure~\ref{fig:tilings}(b)). In the $\xi'$-phase isolated
combinations of a pentagon and a banana-shaped nonagon
(figure~\ref{fig:tilings}(c)) are observed, along which the orientation of the
hexagons is inverted and which are able to move by
flips. \cite{itapdb:Klein1996} therefore have termed them {\em phason-lines},
following the notation of related defects in quasicrystals. In bending
experiments these phason-lines order into periodically aligned {\em
phason-planes} with $1,2, \ldots, n$ rows of hexagons in between
(figure~\ref{fig:tilings}(d) and (e)), forming periodic superstructures. Here
we propose to name these phases $\xi'_{2}$-, $\xi'_{3}$-, $\ldots$,
$\xi'_{n+1}$-phases, the reason for the index-shift being given later. Most
observed is the $\xi'_{2}$-phase, which is also known as $\Psi$-phase
\cite[]{itapdb:Klein1997c}.

The $\xi'_{n}$-phases can be considered striped defect lattices. The stripes,
which are the phason-planes, can bend and move, varying their distances
\cite[]{itapdb:Beraha1997}. \cite{itapdb:Klein1999} have observed dislocations
in the stripe pattern and have called them {\em metadislocations}. They are
the characteristic textures of partial dislocations in the basic tiling.

In this article we are developing a simplified projection formalism to
describe all the $\Xi$-phases. It is derived from the hyperspace method of the
quasiperiodic phases. As a consequence, phasonic degrees of freedom can exist
in the $\Xi$-phases which allow movements of the phason-planes. We determine
those hyperlattice Burgers vectors which label the partial dislocations of
lowest energies and relate them to the Burgers vectors of the
metadislocations. Indeed, these are the only ones which are observed.  The
charming feature of our formalism is that the simplest model is working in
three space so that all steps are easily imaginable.

\section{Geometrical models for i-Al-Pd-Mn and its approximants}

The atomic positions for a quasicrystal can be described by decorating the
lattice points of a periodic hypercrystal with atomic surfaces and marking the
points where these are intersected by a planar cut space (cut-method). An
equivalent method is the strip-method \cite[]{itapdb:Katz1986a}. Approximants
are constructed by using inclined cut spaces \cite[]{itapdb:Duneau1994}. For a
detailed explanation see also \cite{itapdb:Gratias1995}.

\subsection{Sixdimensional cut-method for the i-phase and approximants}

A six-dimensional hyperspace with orthogonal basis vectors $\ve{e}^{6D}_{1}$,
$\ldots$, $\ve{e}^{6D}_{6}$ of length $l^{6D}=0.645$~nm was used by
\cite{itapdb:Katz1994d} to model the i-phase of Al-Pd-Mn. The hyperspace was
decorated with three different atomic surfaces positioned on different nodes
in a face-centred lattice, namely
\begin{itemize}
\item even nodes: n$_{0}=\{(z_{1},\ldots,z_{6}),z_{i}\in\mathbbm{Z}
  \mid\sum_{i}z_{i}=0\mod 2\}$,
\item odd nodes: n$_{1}=\{(z_{1},\ldots,z_{6}),z_{i}\in\mathbbm{Z}
  \mid\sum_{i}z_{i}=1\mod 2\}$,
\item even body-centre nodes: bc$_{0}=\{(z_{1}+\frac{1}{2},\ldots,
  z_{6}+\frac{1}{2}),z_{i}\in\mathbbm{Z}\mid\sum_{i}z_{i}=0\mod 2\}$.
\end{itemize}
The cut space is spanned by three vectors, whose components in the above basis
are ($\tau=\frac{1}{2}(\sqrt{5}+1)$ is the golden mean):
\begin{equation}
\ve{a}^{6D}_{i}=(1,\tau,0,-1,\tau,0),~
\ve{b}^{6D}_{i}=(\tau,0,1,\tau,0,-1),~
\ve{c}^{6D}_{i}=(0,1,\tau,0,-1,\tau).
\end{equation}

\cite{itapdb:Beraha1997} applied the cut-method for the three-dimensional
atomistic description of the phases $\xi$ and $\xi'$ as approximants of the
icosahedral phase. In order to position each single atom correctly, they had
to modify the atomic surfaces slightly. It turned out that the centre volumes
of the n$_{0}$ atomic surfaces, which lie again on a six-dimensional
face-centred lattice, correspond to the cluster centre atoms. This way the
model is reduced to three-dimensional tilings for the cluster positions only.

For the $\xi'$-phase Beraha et al.\ have derived the vectors that span a unit
cell of the inclined cut space. We add the vectors for the inclined cut spaces
of the $\xi$-phase and the $\xi'_{n}$-phases:
\begin{subequations}
\begin{eqnarray}
&&\ve{a}^{6D}_{\xi}=(0,1,1,-1,0,1),~\ve{b}^{6D}_{\xi}=\frac{1}{2}(5,1,1,1,1,-1),~
\ve{c}^{6D}_{\xi}=(0,0,1,1,-1,1)\\
&&\ve{a}^{6D}_{\xi'}=(0,2,1,-2,1,2),~\ve{b}^{6D}_{\xi'}=\ve{b}^{6D}_{\xi},~
\ve{c}^{6D}_{\xi'}=(0,0,1,1,-1,1)\\
&&\ve{a}^{6D}_{\xi'_{n}}=(0,2,1,-2,1,2),~\ve{b}^{6D}_{\xi'_{n}}=\ve{b}^{6D}_{\xi},~
\ve{c}^{6D}_{\xi'_{n}}=(0,0,2n+1,2n,-2n-1,2n)
\end{eqnarray}
\end{subequations}
The $\ve{b}^{6D}_{\xi}$-vector is the same for all the phases, because it
marks the periodicity in the tenfold direction of the d-phase, which coincides
with the column line of the Mackay-type clusters. By projecting in direction
of $\ve{b}^{6D}_{\xi}$, two-dimensional tilings like those in
figure~\ref{fig:tilings} can be obtained. The edge length of the tiles is
$t^{6D}=\frac{1}{5}\sqrt{10}\sqrt{\tau+2}\,l^{6D}=0.78$~nm. It can be
calculated by projecting all connection vectors
$\ve{e}_{i}^{6D}+\ve{e}_{j}^{6D}$ of neighbouring n$_{0}$-sites onto the
tiling plane. The shortest projections have length $t^{6D}$.

Since the six vectors $\ve{a}^{6D}_{\xi}$, $\ve{c}^{6D}_{\xi}$,
$\ve{a}^{6D}_{\xi'}$, $\ve{c}^{6D}_{\xi'}$, $\ve{a}^{6D}_{\xi'_{n}}$ and
$\ve{c}^{6D}_{\xi'_{n}}=2n\ve{c}^{6D}_{\xi'}-\ve{a}^{6D}_{\xi'}+2\ve{a}^{6D}_{\xi}$
lie in a three-dimensional subspace of the six-dimensional hyperspace, a
description of the tilings in a three-dimensional hyperspace is possible by
the cut-method. The model can be simplified by using a simple cubic lattice in
three-space, as will be shown in detail in section
\ref{sec:threedimensional}. We can arrive at this conclusion directly after
substituting the tiling of hexagons, pentagons and nonagons by a tiling of
rhombs.

\subsection{Rhombic substitution tiling}

The new tiles are the thin and thick Penrose rhombs which in general are
spanned by the vectors of a regular five-star. The interior angles of the
tiles are multiples of $36^{\circ}$.

\begin{figure}\begin{center}
\includegraphics[width=16cm]{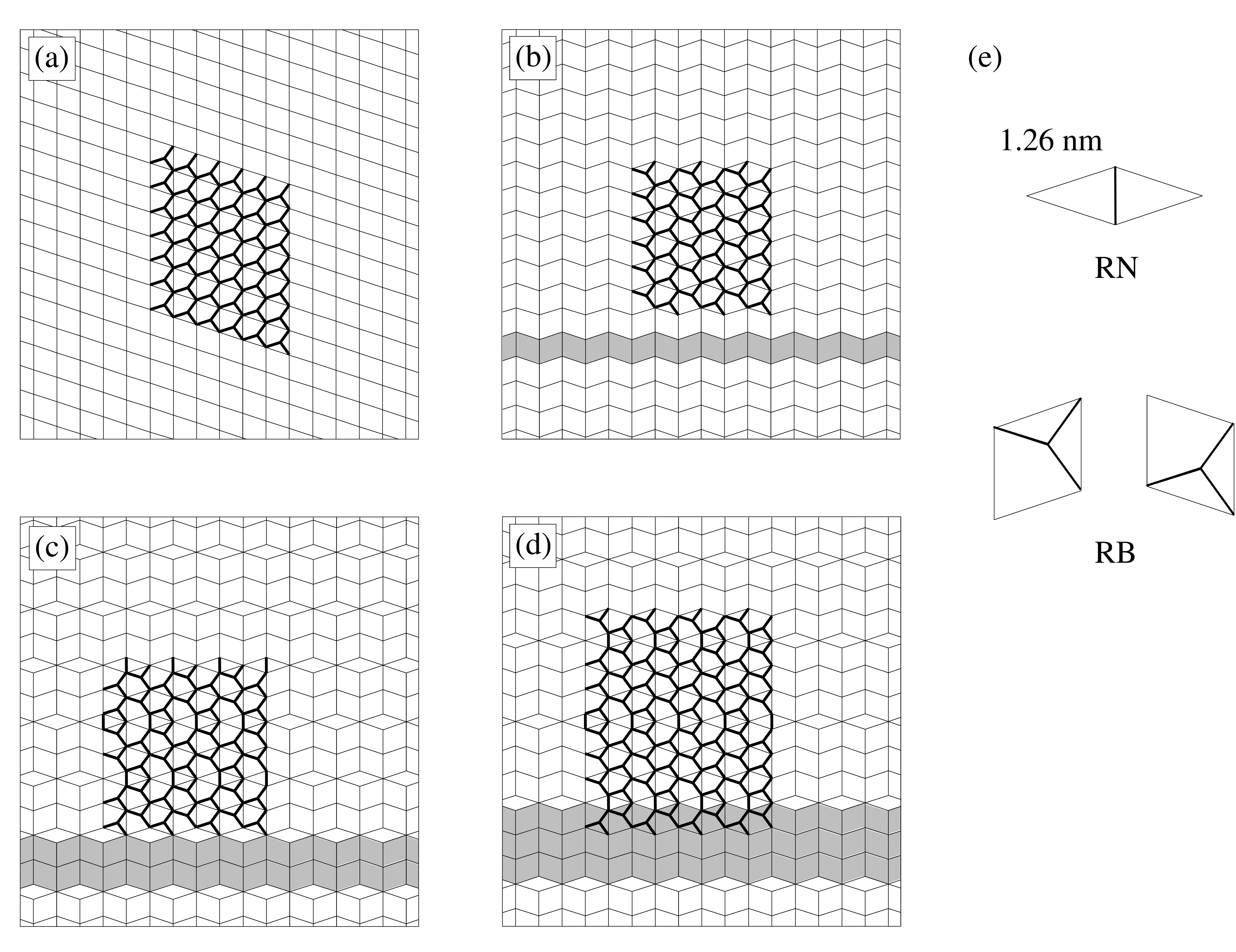}
  \caption{New tilings can be created by substituting the original tiles with
    thin Penrose rhombs RN, called phason-lines, and thick Penrose rhombs
  RB. The $\xi$-phase~(a) and the $\xi'$-phase~(b) are built only with the
  thick rhombs, while for the $\xi'_{2}$-phase~(c) and the
  $\xi'_{3}$-phase~(d) also the thin rhombs are needed. Between two
  phason-planes in the $\xi'_{n}$-phase there are $n$ rows of alternating
  thick rhombs (marked grey). The new tiles are shown in~(e). The edge length
  of the tiles is: $t^{5D}=1.26$~nm. \label{fig:tilings2}}
\end{center}\end{figure}

As shown in figure~\ref{fig:tilings2}, a hexagon of the original tiling is
substituted by a thick rhomb, and a nonagon/pentagon combination is
substituted by a combination of a thin and a thick rhomb.\footnote{A similar
  tiling has been presented by Klein et al. (1996). They used thick rhombs to
  model the $\xi$- and $\xi'$-phase and small hexagons for the
  phason-lines. However no description in hyperspace was presented, explaining
  the arrangement of the tiles.}
Therefore we will refer to the thin rhombs as phason-lines. A row of hexagons
in staggered orientation is substituted by a row of alternating thick rhombs,
and a phason-plane is substituted by a combination of a row of alternating
thick rhombs and a row of thin rhombs, which again will be called
phason-plane. Hence the number of rows of thick rhombs between two
phason-planes in the new tiling for the $\xi'_{n}$-phase is exactly $n$.

Thus the tilings for the column positions of the $\Xi$-phases emerge as
approximants of the Penrose tiling. However, only rhombs of three orientations
show up. To span these only three prongs of the five-star are required.
This is another argument why we can apply a projection formalism in a
three-dimensional hyperspace which was mentioned above and is elaborated
further below.

To model the lattices of all $\Xi$-phases, their phasonic degrees of freedom
and their metadislocations we found it suitable to resort to a modified cut and
projection method. To our knowledge it has not been applied yet in the
literature. It makes use of {\em atomic hypervolumes} and requires a short
section.

\subsection{Atomic hypervolumes for a geometrical description in hyperspace}

\begin{figure}\begin{center}
\includegraphics[width=10cm]{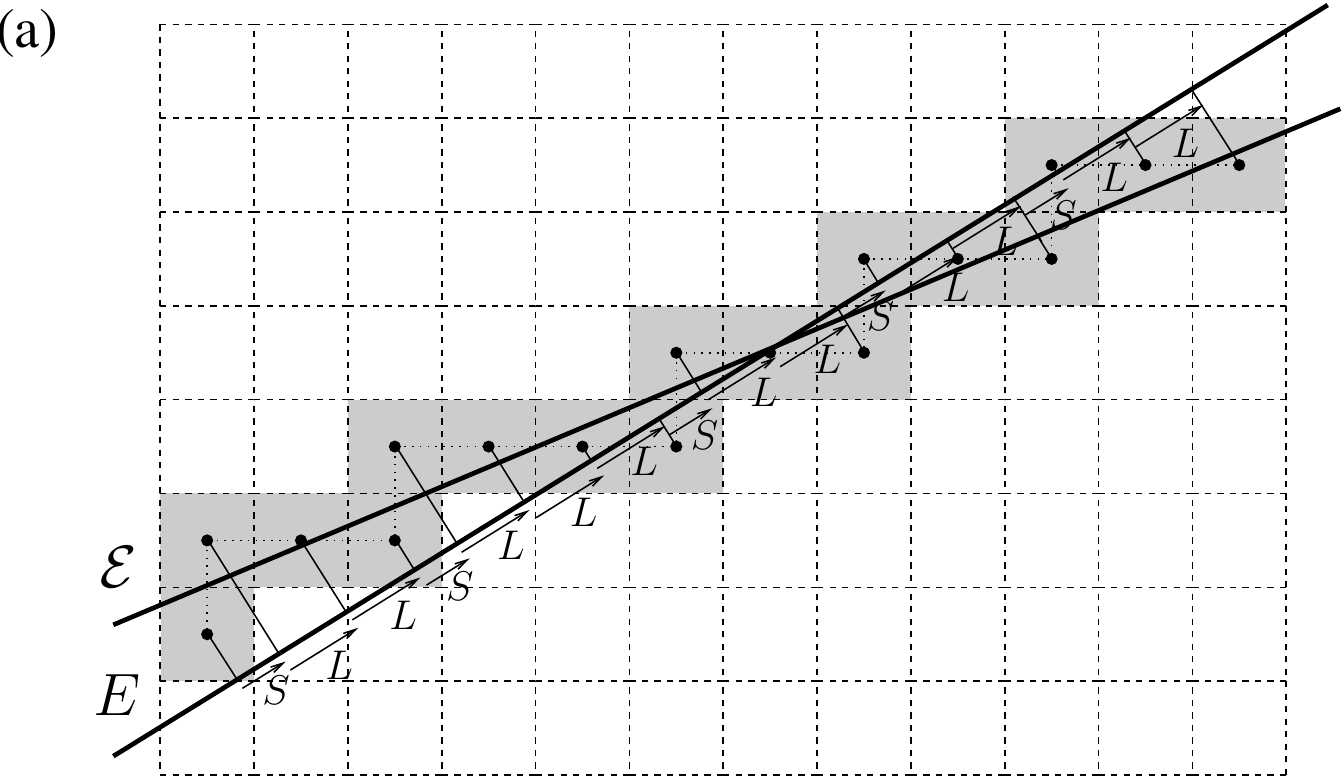}\vspace{1cm}
\includegraphics[width=10cm]{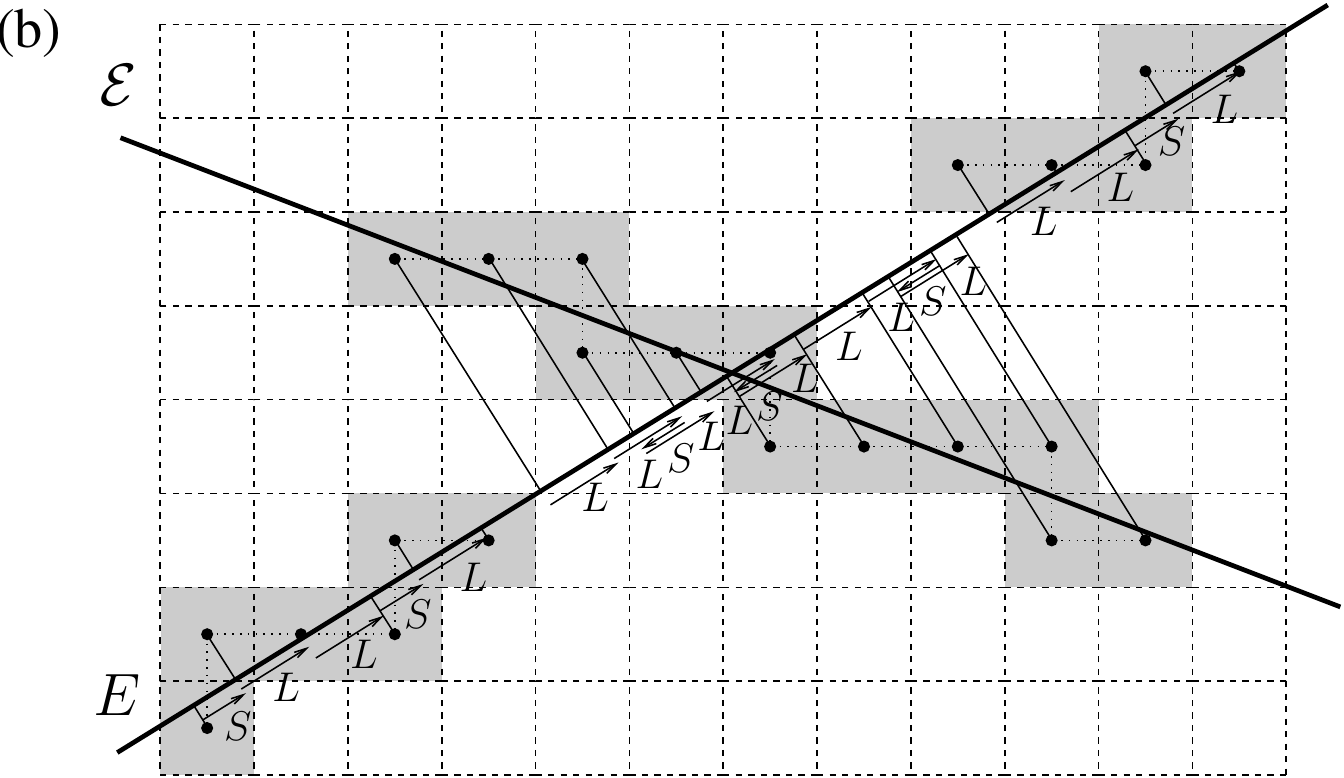}
  \caption{(a)~Construction of an approximant of the Fibonacci-chain by the
    method of atomic volumes. A cut line $\mathcal{E}$ and a projection line
  $E$ are required (see text). (b)~If the slope of $\mathcal{E}$ is negative,
  the projected tiles overlap. This case is geometrically not allowed. In the
  bottom left and top right corner a small part of the quasiperiodic
  Fibonacci-chain is drawn. \label{fig:fibonacci}}
\end{center}\end{figure}

The method will be explained by the example of the well known one-dimensional
Fibonacci-chain: The two-dimensional hyperspace is partitioned into equal unit
cells, called atomic volumes (in general atomic hypervolumes) as shown in
figure~\ref{fig:fibonacci}(a).\footnote{As a generalisation several
  overlapping atomic volumes could be used, as long as the number of times a
  point is covered is constant for the hyperspace. This way the closeness
  condition \cite[]{itapdb:Frenkel1986a} of the original cut-method is
  automatically fulfilled. But we will only make use of the method of atomic
  volumes for the canonical case of a hyper-cubic lattice $\mathbbm{Z}^n$ and
  the unit cells as atomic volumes.}
For the construction of an approximant, two different lines are needed: the
cut line $\mathcal{E}$ (in general cut plane or cut space) and the projection
line $E$ (in general projection plane or projection space, also named physical
space). Those cells that are cut by $\mathcal{E}$ are selected (marked
grey). The middle point of those cells is projected onto the projection
line.\footnote{Any other choice is possible too, as long as it is the same
  point for each cell, since a different selection only leads to a global
  translation of the tiling.}
The projection leads to two different intervals on $E$, either a small one
($S$), when the neighbouring selected cells meet vertically, or a large one
($L$), when they meet horizontally, forming the tiles of the Fibonacci-chain.

The orientation of $E$ determines the shape of the tiles. In the example of
the Fibonacci-chain, it determines the ratio of $S$ to $L$. In contrast, the
orientation of $\mathcal{E}$ fixes the arrangement of the tiles. Since
different approximants of the same quasiperiodic phase are built from the same
tiles, but with different arrangement of the tiles, the only difference in the
construction is the varying orientation of $\mathcal{E}$. A rational slope of
the cut line leads to a periodic tiling, while an irrational slope generates a
quasiperiodic tiling. The special case of $E$ and $\mathcal{E}$ being
identical is the original cut-method as mentioned above.

A geometrical restriction on the orientation of $\mathcal{E}$ is imposed by
the fact, that projected tiles must not overlap. In the example of the
Fibonacci-chain this means, that the slope of $\mathcal{E}$ must be
positive. For a counterexample when it is not fulfilled see
figure~\ref{fig:fibonacci} (b). With this method we now construct the tilings
of the cluster projections.

In the original cut-method the shape of the atomic surfaces depends on the
orientation of the cut space. For the Fibonacci-chain the atomic surfaces are
the projections of the unit cells on the orthogonal complement of
$\mathcal{E}$. Because we will deal with cut spaces with spatially varying
orientation in this paper, we would need a varying shape of the atomic
surfaces as well. The method of atomic hypervolumes avoids these problems, as
the atomic hypervolumes themselves do not vary.

\subsection{Three- and five-dimensional hyperspace for approximants of Al-Pd-Mn}
\label{sec:threedimensional}

\begin{figure}\begin{center}
\includegraphics[width=11cm]{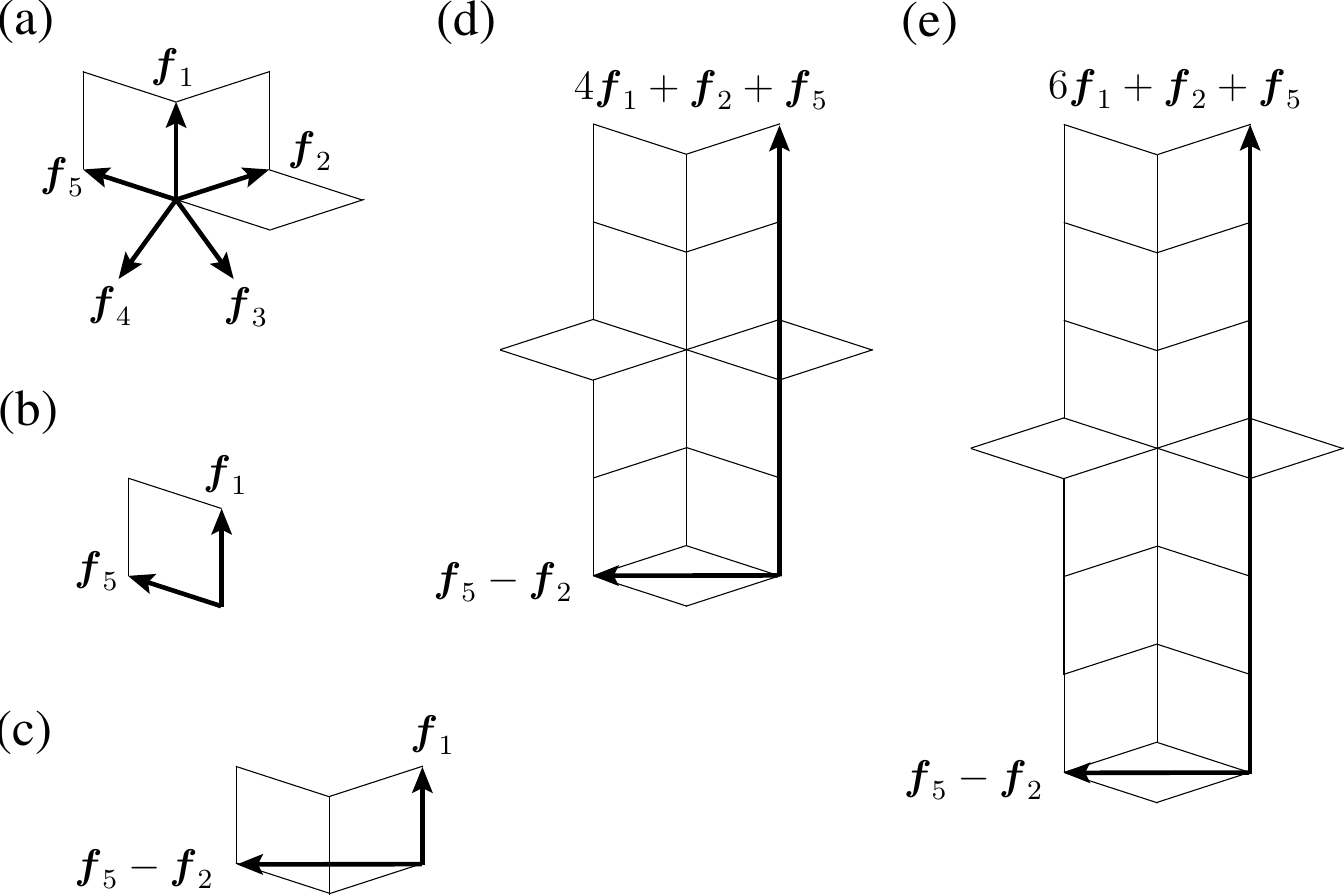}
  \caption{(a)~The projected basis vectors $\ve{f}_{i}$ lie on a regular
    five-star. The three rhombs of the tilings in figure \ref{fig:tilings2}
    can be constructed with $\ve{f}_{1}$, $\ve{f}_{2}$ and $\ve{f}_{5}$. Unit
    cells of approximants and their bases are shown: the $\xi$-phase~(b), the
    $\xi'$-phase~(c), the $\xi'_{2}$-phase~(d) and the
    $\xi'_{3}$-phase~(e). \label{fig:fivestar}}
\end{center}\end{figure}

Because the rhombic tilings for the $\Xi$-phases are approximants of the
Penrose tiling they can be described in a five-dimensional hyperspace with basis
$\{\ve{e}^{5D}_{1},\ldots,\ve{e}^{5D}_{5}\}$, the hyper-cubic lattice
$\mathbbm{Z}^{5}$ and the unit cells as atomic hypervolumes. The projection
plane of the Penrose tiling is spanned by the vectors
\begin{eqnarray}
&&\ve{a}^{5D}_{p}=\left(\sin(2\pi\tfrac{0}{5}), \sin(2\pi\tfrac{1}{5}),
\sin(2\pi\tfrac{2}{5}), \sin(2\pi\tfrac{3}{5}),
\sin(2\pi\tfrac{4}{5})\right),\nonumber\\
&&\ve{c}^{5D}_{p}=\left(\cos(2\pi\tfrac{0}{5}), \cos(2\pi\tfrac{1}{5}),
\cos(2\pi\tfrac{2}{5}), \cos(2\pi\tfrac{3}{5}), \cos(2\pi\tfrac{4}{5})\right)
\end{eqnarray}
and the projection matrix is obtained by writing these vectors normalised in
the rows:
\begin{equation}
\pi^{5D}_{\parallel}=\frac{1}{5}\sqrt{10}\left(\begin{array}{ccccc}
\sin(2\pi\frac{0}{5}) & \sin(2\pi\frac{1}{5}) & \sin(2\pi\frac{2}{5}) &
\sin(2\pi\frac{3}{5}) & \sin(2\pi\frac{4}{5})\\
\cos(2\pi\frac{0}{5}) & \cos(2\pi\frac{1}{5}) & \cos(2\pi\frac{2}{5}) &
\cos(2\pi\frac{3}{5}) & \cos(2\pi\frac{4}{5}) \\
\end{array}\right).
\end{equation}
The projections $\ve{f}_{i}=\pi^{5D}_{\parallel}\ve{e}^{5D}_{i}$ of the
five-dimensional basis vectors give a regular five-star as drawn in
figure~\ref{fig:fivestar}(a). The rhomb tiles are the projection of
two-dimensional faces of the five-dimensional unit cells. Each is spanned by
two vectors of the five-star.

The edge length of the rhomb tiles can be calculated from the edge length of
the pentagon/hexagon/nonagon-tiles as shown in figure \ref{fig:tilings2}(e):
$t^{6D}=2\cos(72^{\circ})\,t^{5D}=\tau\,t^{5D}$. This gives:
$t^{5D}=\frac{1}{5}\sqrt{10}\,\tau\sqrt{\tau+2}l^{6D}=1.26$~nm, leading to the
length of the basis vectors in the five-dimensional hyperspace:
$l^{5D}=\tau\sqrt{\tau+2}\,l^{6D}=1.99$~nm.

A basis $\{\ve{a}^{5D},\ve{c}^{5D}\}$ for the cut plane of a periodic
approximant can be determined with a basis of the unit cell of the
tiling. This is done in figures~\ref{fig:fivestar}(b)-(e) for various
approximant phases:
\begin{subequations}
\label{eq:basis5d}
\begin{eqnarray}
&&\ve{a}^{5D}_{\xi}=(0,0,0,0,1),~\ve{c}^{5D}_{\xi}=(1,0,0,0,0),\\
&&\ve{a}^{5D}_{\xi'}=(0,-1,0,0,1),~\ve{c}^{5D}_{\xi'}=(1,0,0,0,0),\\
&&\ve{a}^{5D}_{\xi'_{n}}=(0,-1,0,0,1),~\ve{c}^{5D}_{\xi'_{n}}=(2n,1,0,0,1).
\end{eqnarray}
\end{subequations}

We now want to derive the transition matrix $T^{6D}_{5D}$ from the five- to
the six-dimensional hyperspace which maps the basis vectors $\ve{a}^{5D}$ and
$\ve{c}^{5D}$ on $\ve{a}^{6D}$ and $\ve{c}^{6D}$. The first column of
$T^{6D}_{5D}$ is fixed by the relation
$\ve{c}^{6D}_{\xi}=T^{6D}_{5D}\ve{c}^{5D}_{\xi}$. The other columns follow
from symmetry considerations. The fivefold rotation of the hyperspaces that
fixes the projection plane is for the five- and six-dimensional model:
\begin{equation}
\label{eq:rotations}
R^{5D}=\left(\begin{array}{ccccc}
0 & 1 & 0 & 0 & 0\\
0 & 0 & 1 & 0 & 0\\
0 & 0 & 0 & 1 & 0\\
0 & 0 & 0 & 0 & 1\\
1 & 0 & 0 & 0 & 0\\
\end{array}\right),\qquad
R^{6D}=\left(\begin{array}{cccccc}
1 & 0 & 0 & 0 & 0 & 0\\
0 & 0 & 1 & 0 & 0 & 0\\
0 & 0 & 0 & 1 & 0 & 0\\
0 & 0 & 0 & 0 & 0 &-1\\
0 & 1 & 0 & 0 & 0 & 0\\
0 & 0 & 0 & 0 &-1 & 0\\
\end{array}\right).
\end{equation}
Observe, that $R^{6D}$ does not change the vector $\ve{b}^{6D}_{\xi}$. The
symmetry operation in the two hyperspace must match:
$R^{6D}T^{6D}_{5D}=T^{6D}_{5D}R^{5D}$. This determines the $(i+1)$-th column
of $T^{6D}_{5D}$ via
\begin{equation}
\ve{c}^{6D}_{\xi}=[R^{6D}]^{-i}\circ T^{6D}_{5D}\circ
[R^{5D}]^{i}\,\ve{c}^{5D}_{\xi}
\end{equation}
resulting in:
\begin{equation}
\label{eq:transition5D6D}
T_{5D}^{6D}=\left(\begin{array}{ccccc}
 0 &  0 &  0 &  0 &  0\\
 0 & -1 & -1 &  1 &  1\\
 1 &  0 & -1 & -1 &  1\\
 1 &  1 &  0 & -1 & -1\\
-1 & -1 &  1 &  1 &  0\\
 1 & -1 & -1 &  0 &  1\\
\end{array}\right).
\end{equation}

After fixing a coordinate system for the atomic hypervolumes in the
five-dimensional hyperspace, the approximant phases can appear in five
different orientations corresponding to a cyclic permutation of the basis
vectors. Each orientation of the phases $\xi'$ and $\xi'_{n}$ is characterised
by the orientation of the thin rhombs. By fixing their orientation, a
lower-dimensional hyperspace can be used. Since the third and fourth
components of the basis vectors in (\ref{eq:basis5d}) vanish, a description in
a {\em three-dimensional} hyperspace with cubic lattice $\mathbbm{Z}^{3}$
(lattice parameter $l^{5D}=l^{3D}$) and the unit cell as atomic volume is
possible by omitting these components. The transition matrix is simply:
\begin{equation}
\label{eq:transition3D5D}
T_{3D}^{5D}=\left(\begin{array}{ccc}
 1 &  0 &  0\\
 0 &  1 &  0\\
 0 &  0 &  0\\
 0 &  0 &  0\\
 0 &  0 &  1\\
\end{array}\right).
\end{equation}
For the $\Xi$-phases this three-dimensional model leads to the same tilings as
the five-dimensional model, but only one orientation of the thin rhombs and
two orientations of the thick rhombs can show up. The projection matrix
$\pi^{3D}_{\parallel}$ is derived from $\pi^{5D}_{\parallel}$ again by
omitting the third and fourth components:\footnote{This projection is not
  orthogonal, while the projection $\pi^{5D}_{\parallel}$ is orthogonal. That
  is a consequence of the omission of two hyperspace dimensions.}
\begin{eqnarray}
\pi^{3D}_{\parallel}&=&\frac{1}{5}\sqrt{10}\left(\begin{array}{ccccc}
\sin(2\pi\frac{0}{5}) & \sin(2\pi\frac{1}{5}) & \sin(2\pi\frac{4}{5})\\
\cos(2\pi\frac{0}{5}) & \cos(2\pi\frac{1}{5}) & \cos(2\pi\frac{4}{5})\\
\end{array}\right)\notag\\
&=&\frac{1}{5}\sqrt{10}\left(\begin{array}{ccccc}
0 & \frac{1}{2}\sqrt{2+\tau} & -\frac{1}{2}\sqrt{2+\tau}\\
1 & \frac{1}{2}\tau^{-1} & \frac{1}{2}\tau^{-1}\\
\end{array}\right).
\end{eqnarray}

Bases for the cut planes in the three-dimensional model are given by
\begin{subequations}
\begin{eqnarray}
&&\ve{a}^{3D}_{\xi}=(0,0,1),~\ve{c}^{3D}_{\xi}=(1,0,0),\\
&&\ve{a}^{3D}_{\xi'}=(0,-1,1),~\ve{c}^{3D}_{\xi'}=(1,0,0),\\
&&\ve{a}^{3D}_{\xi'_{n}}=(0,-1,1),~\ve{c}^{3D}_{\xi'_{n}}=(2n,1,1).
\end{eqnarray}
\end{subequations}

\subsection{Relationships of the hyperspaces}

So far the three- and five-dimensional hyperspace have been introduced only
phenomenologically for the description of tilings of the approximant
phases. Now we want to show how they are related to the original
six-dimensional hyperspace. This is meant as a clarification of hyperspace
geometry. Mathematically, the formation of the approximants from the
icosahedral quasicrystal proceeds in two steps:
\begin{enumerate}
\item[(i)] The configuration in direction of a fivefold rotation axis $\ve{b}$
  is changed, while the configuration perpendicular to this direction in the
  plane spanned by $\ve{a}$ and $\ve{c}$ is unaltered, leading to a
  pentagonal or decagonal quasicrystal. This step cannot be described simply
  by a reorientation of the cutplane. Additionally a relaxation of the atoms
  in direction of $\ve{b}$ is necessary as shown by \cite{itapdb:Beraha1997}.
\item[(ii)] The configuration in the plane spanned by $\ve{a}$ and $\ve{c}$ is
  changed, while the configuration in direction of $\ve{b}$ is unaltered. This
  step is accurately described by a reorientation of the cutplane as used in
  this paper.
\end{enumerate}

In step (i) a fivefold symmetry of the icosahedral quasicrystal is
conserved. The fivefold rotation $R^{6D}$ (equation (\ref{eq:rotations}))
operates trivially on a two-dimensional subspace $U_{1}$ spanned by
$(1,0,0,0,0,0)$, $(0,1,1,1,1,-1)$ and as a true rotation on the
four-dimensional orthogonal complement $U_{2}$ spanned by $(0,0,1,1,-1,1)$,
$(0,-1,0,1,-1,-1)$, $(0,-1,-1,0,1,-1)$, $(0,1,-1,-1,1,0)$. So for
quasicrystals and approximants the lattice vector $\ve{b}$ is projected from
$U_{1}$, and the vectors $\ve{a}$ and $\ve{c}$ are projected from $U_{2}$.

\begin{figure}\begin{center}
\includegraphics[width=14cm]{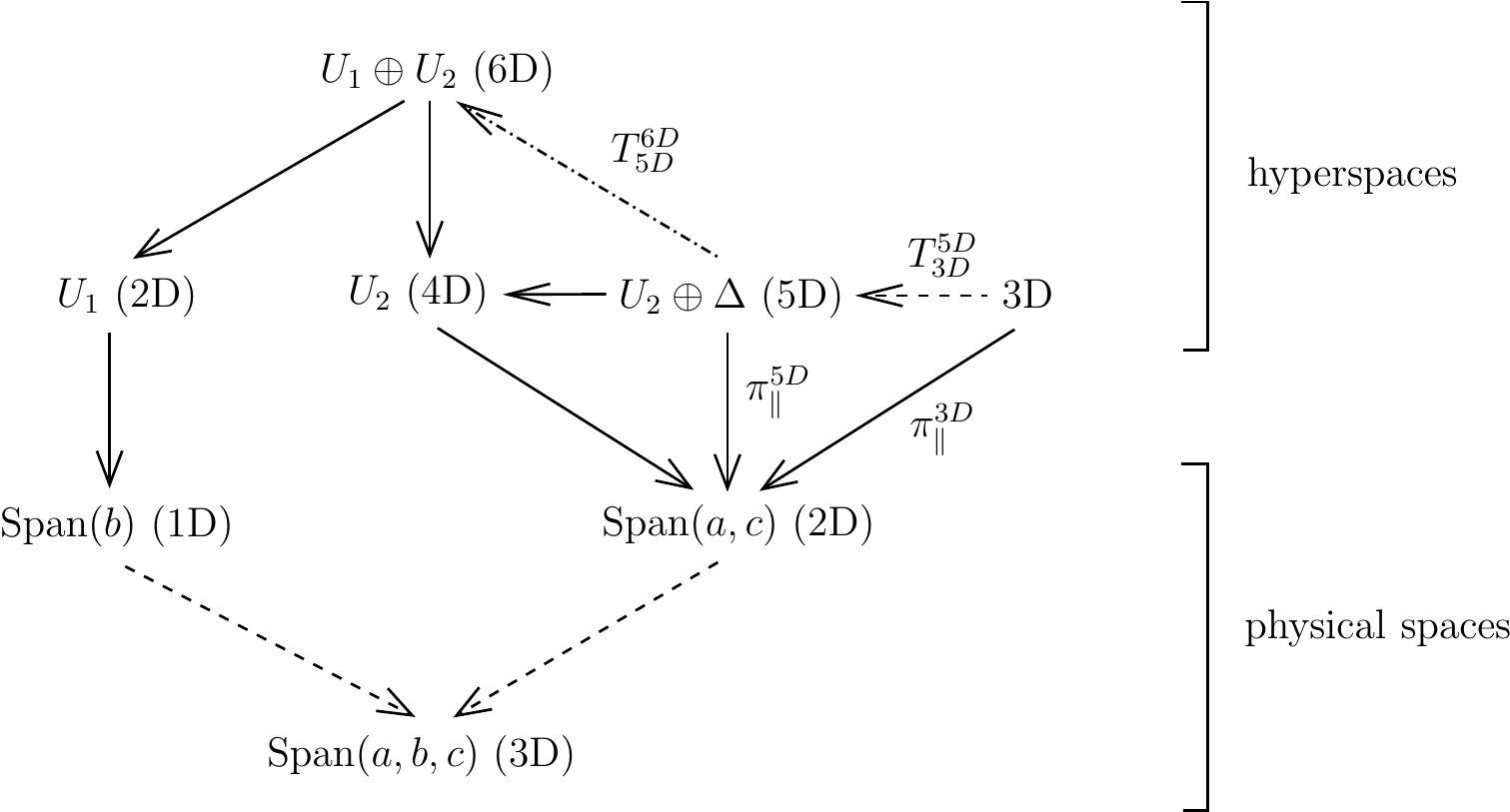}
  \caption{Relationship of the hyperspaces and the physical spaces. The
    five-dimensional space is built on top of the four-dimensional space
    $U_{2}$ perpendicular to the fivefold two-dimensional plane $U_{1}$. It is
    not a (full) subspace of the original six-dimensional space. Solid arrows
    indicate projections (subspace relationship). Dashed arrows indicate
    inclusions. $T_{5D}^{6D}|_{U_{2}}$ is an inclusion and
    $T_{5D}^{6D}|_{\Delta}=0$ the zero mapping. \label{fig:hyperspaces}}
\end{center}\end{figure}

The restriction of $\mathbbm{Z}_{6}$ onto $U_{2}$ leads to a non-cubic
four-dimensional lattice (the $\mathbbm{A}_{4}$ root lattice) that could in
principle be used for the construction of the Penrose tiling and its
approximant tilings. By adding a one-dimensional complementary space $\Delta$,
these tilings can be constructed from the $\mathbbm{Z}^{5}$-lattice. This is
well-known and the usual way to describe the Penrose lattice. The relation of
the five-dimensional hyperspace to the three-dimensional hyperspace is
described in section \ref{sec:threedimensional}. Figure \ref{fig:hyperspaces}
summarizes all the hyperspace relationships.

\section{Phasonic degrees of freedom}
\label{sec:phasonfreedom}

Besides standard elastic ({\em phononic}) degrees of freedom, quasicrystals
have additional degrees of freedom \cite[]{itapdb:Socolar1986b}, which
originate from the fact that the cut space is embedded in hyperspace. Local
{\em excitations} of these so-called {\em phasonic} degrees of freedom,
correspond to continuous displacements of the cut space along the direction
orthogonal to it in hyperspace. The direction of the excitation is understood
as the direction of the displacement.

The number of the phasonic degrees of freedom for a $n$-dimensional hyperspace
and a $d$-dimensional cut space is equal to $n-d$. As will be shown, phasonic
degrees of freedom are also possible for approximant phases, although then
they are not any more continuous degrees of freedom. For convenience we
nonetheless continue using this notation.

\subsection{Restriction to excitable degrees of freedom}

\begin{figure}\begin{center}
\includegraphics[width=10cm]{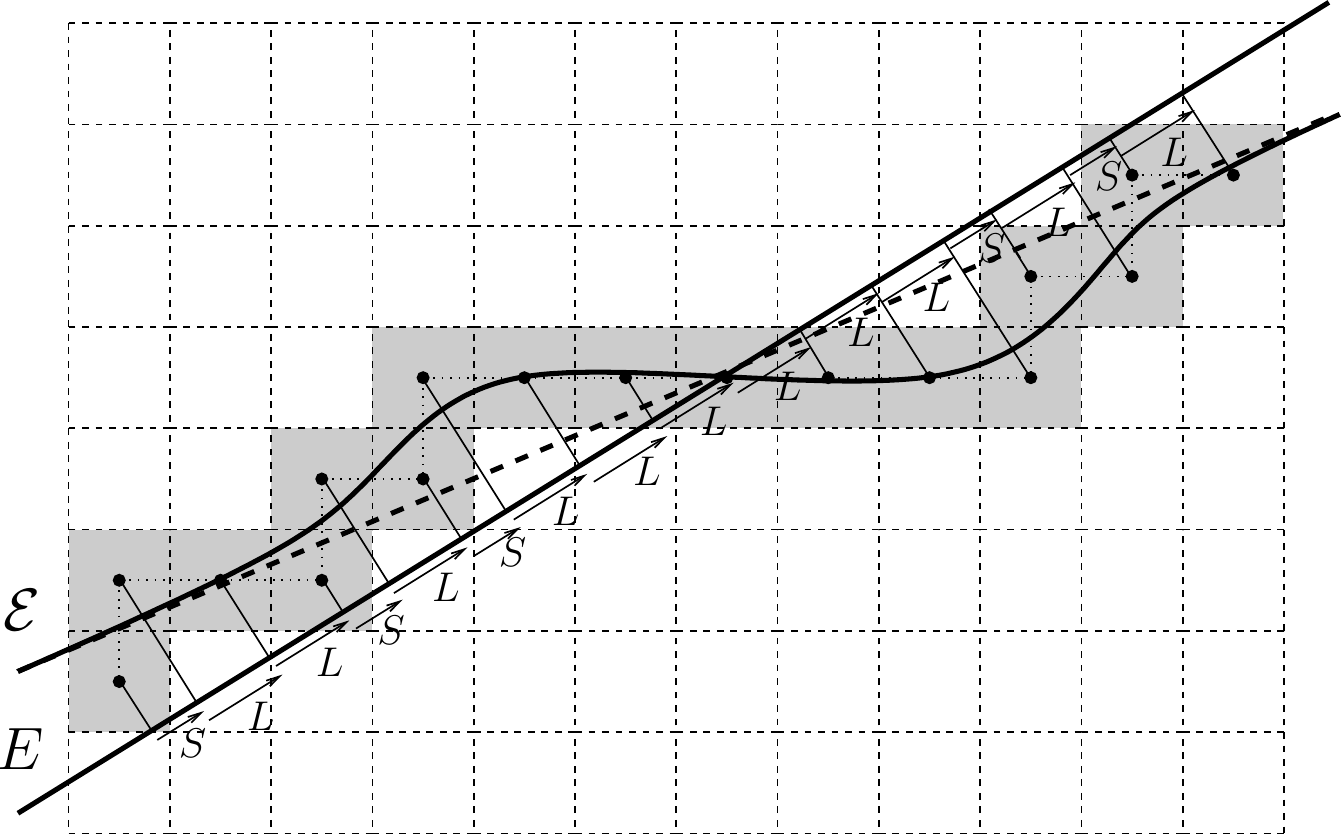}
  \caption{Local excitation of the phasonic degree of freedom for an
    approximant of the Fibonacci-chain. The excitation rearranges the sequence
    of the tiles. \label{fig:fibonacci2}}
\end{center}\end{figure}

In the example of the Fibonacci-chain, a local excitation of the phasonic
degree of freedom is shown in figure~\ref{fig:fibonacci2}. Because of the
displacement the cut line $\mathcal{E}$ now selects different atomic
volumes. In the tiling this leads to a local rearrangement of the tiles,
called phasonic jumps. In general the excitations are considered small and
distributed over a wide tiling region, so that the orientational deviation
from the planar cut space is small.

A special case has to be examined separately: If the cut plane of an
approximant is parallel to boundaries of atomic volumes, it is possible that a
local excitation of phasonic degrees of freedom leads to overlapping tiles.
If, for example in the Fibonacci-chain, the cut line of the approximant has
slope 0, a local excitation always creates regions of the cut line having
positive as well as negative slope. The regions of negative slope lead to
overlapping tiles, as shown in figure~\ref{fig:fibonacci}(b) earlier. In such
a case, accordingly, local displacements of the cut plane are geometrically
not allowed and the phasonic degree of freedom is termed {\em not excitable}.

With regard to the approximants of the 1.6~nm d-(Al-Pd-Mn) in the
five-dimensional model, the phasonic degrees of freedom of the $\Xi$-phases
are not excitable in the directions of $\ve{e}^{5D}_{3}$ and
$\ve{e}^{5D}_{4}$. So for a full description of these phases the
three-dimensional hyperspace with basis $\{\ve{e}^{3D}_{1}=\ve{e}^{5D}_{1},
\ve{e}^{3D}_{2}=\ve{e}^{5D}_{2},\ve{e}^{3D}_{3}= \ve{e}^{5D}_{5}\}$ and only
one phasonic degree of freedom is sufficient.

This last phasonic degree of freedom is excitable in the $\xi'_{n}$-phases,
leading to new phenomena. But it is not excitable in the $\xi$- and the
$\xi'$-phase. (The cut plane of the $\xi'$-phase, for example, is parallel to
the $\ve{e}_{1}^{3D}$-axis). So the phases $\xi$ and $\xi'$ behave like normal
periodic crystals, having only phononic degrees of freedom.

It is a pleasant aspect of the three-dimensional hyperspace that the
construction formalism can easily be visualised. Thus in the next sections we
can present the cut plane for phase boundaries and dislocations as
two-dimensional curved surfaces.

\subsection{Phasonic phase boundaries}

Different approximant phases are characterised by different orientations of
the cut space. As a feature of the phasonic degrees of freedom, a spatially
dependent continuous transformation from one orientation of the cut space to
another is possible, leading to a phasonic phase boundary.

\begin{figure}\begin{center}
\includegraphics[width=7.5cm]{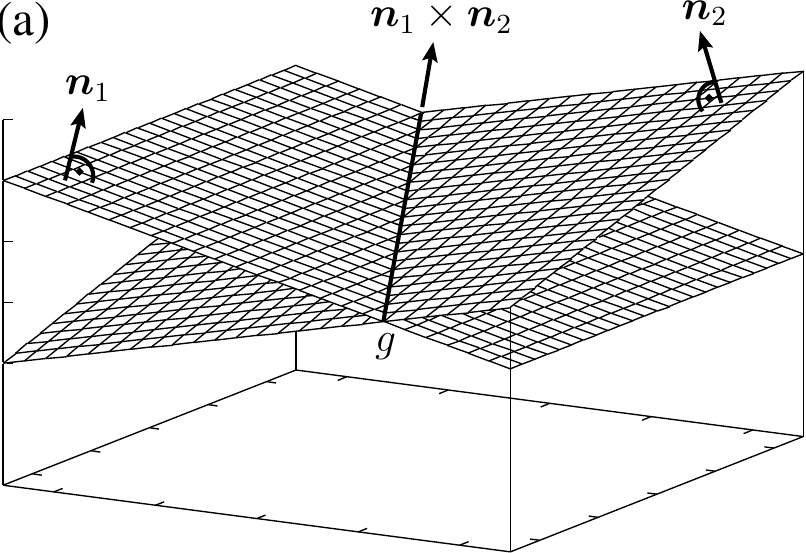}\qquad
\includegraphics[width=7.5cm]{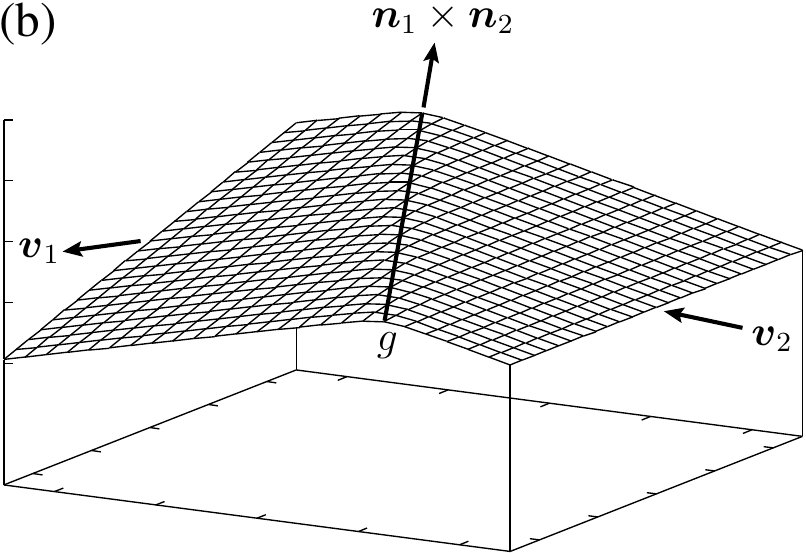}
  \caption{(a)~Two different cut planes with normal vectors $\ve{n}_{1}$ and
    $\ve{n}_{2}$. They have the line $g$ with direction vector
  $\ve{n}_{1}\times\ve{n}_{2}$ in common. (b)~A transition between the two
  phases is realised by adjusting the orientation of the cut plane
  continuously (see text). \label{fig:phaseboundary}}
\end{center}\end{figure}

Let two phases be characterised by the normal vectors $\ve{n}_1$ and
$\ve{n}_2$ of their cut planes $\mathcal{E}_1$ and $\mathcal{E}_2$
(figure~\ref{fig:phaseboundary}(a)). For calculating the cut plane
$\mathcal{E}$ of the phasonic phase boundary we choose the coordinate system
in a way, that the line $g=\mathcal{E}_1\cap\mathcal{E}_2$ with direction
$\ve{n}_1\times\ve{n}_2$ runs through the origin. In the limit far away from
$g$ on the side of phase~1 the new cut plane $\mathcal{E}$ contains the vector
$\ve{v}_{1}=\ve{n}_1\times(\ve{n}_1\times\ve{n}_2)$, while on the other side
$\mathcal{E}$ the vector $\ve{v}_{2}=\ve{n}_2\times(\ve{n}_1\times\ve{n}_2)$.
By using a transition function $f:\mathbbm{R}\rightarrow[0,1]$ with properties
\begin{equation}
\lim_{x\rightarrow-\infty}f(x)=0\text{, }\lim_{x\rightarrow\infty}f(x)=1 \text{
and }f(0)=1/2
\end{equation}
the cut plane for the phase boundary $\mathcal{E}$ as shown in
figure~\ref{fig:phaseboundary}(b) is parametrised by:
\begin{equation}
\mathcal{E}=\left\{s\frac{\ve{n}_1\times\ve{n}_2}{\|\ve{n}_1\times\ve{n}_2\|}+
t\frac{f(t)\ve{v}_{2}+(1-f(t))\ve{v}_{1}}{\|f(t)\ve{v}_{2}+(1-f(t))\ve{v}_{1}\|}
\;;\;s,t\in\mathbbm{R}\right\}.
\end{equation}

\begin{figure}\begin{center}
\includegraphics[width=7.4cm]{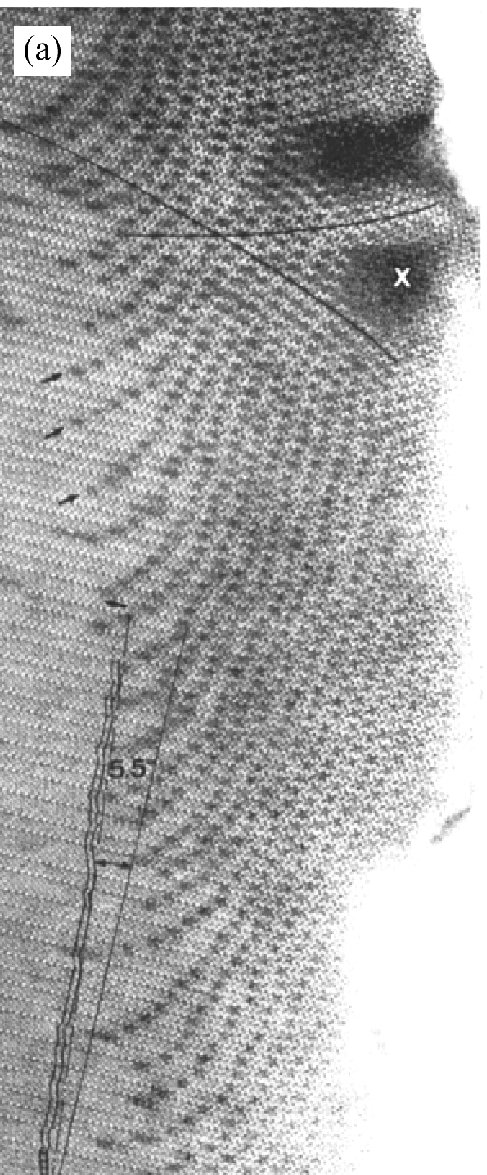}\qquad
\includegraphics[width=7.6cm]{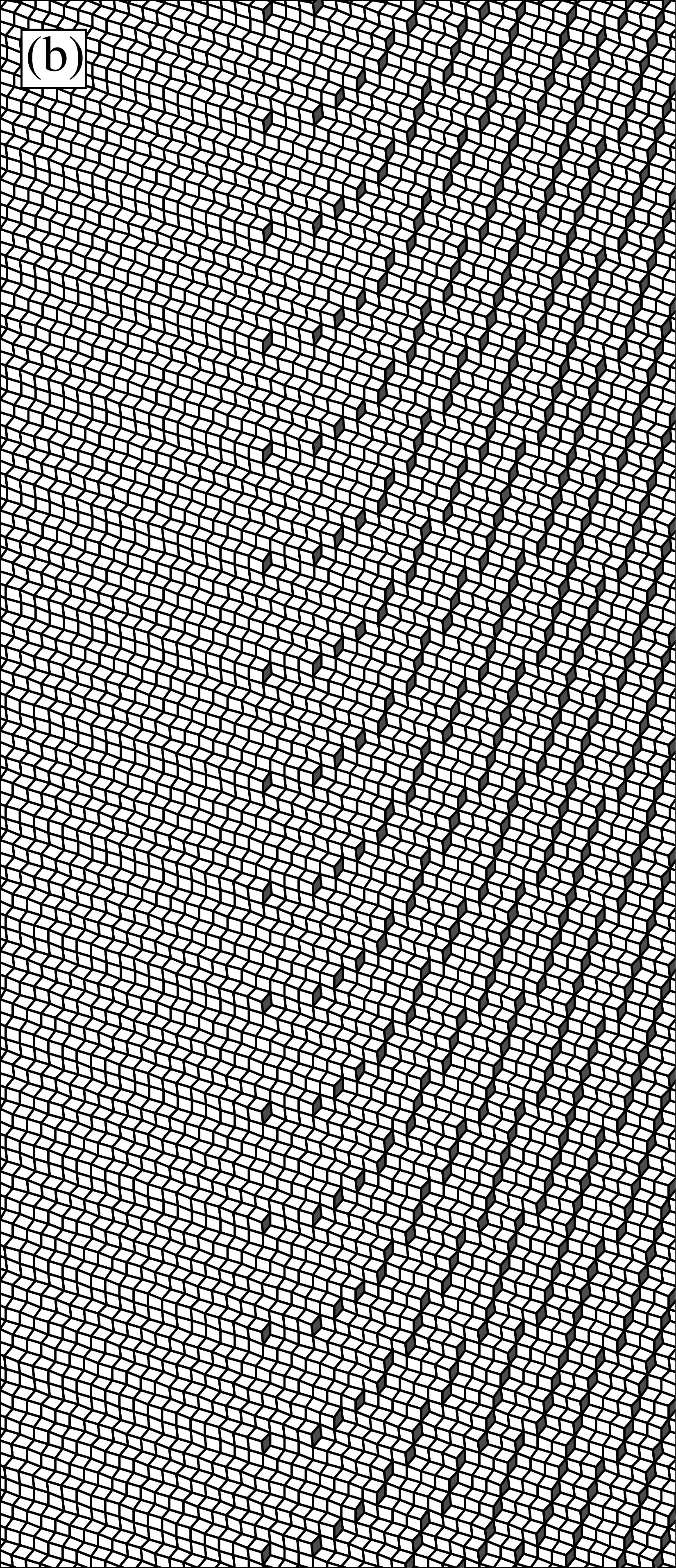}
  \caption{Phasonic phase boundary between the phase $\xi(0.8)$ on the left
    side and the phase $\xi'_{2}$ on the right side. (a)~Transmission electron
  micrograph. The darker spots correspond to phason-lines
  \cite[]{itapdb:Klein1996}. (b)~Calculated tiling for the three-dimensional
  model (see text). \label{fig:phaseboundary2}}
\end{center}\end{figure}

\cite{itapdb:Klein1996} have taken a transmission electron micrograph
of a phasonic phase boundary (figure~\ref{fig:phaseboundary2}(a)),
without interpreting it as such. The phason-lines can be seen as dark
contrasts.  Phase~1 on the left side is one of the phases $\xi(\alpha)$,
$0\leq\alpha\leq 1$ with $\alpha\approx 0.8$, by which we denote an
intermediate phase of a $\xi$- and a $\xi'$-approximant. Its cut plane is
spanned by $\ve{a}_{\xi(\alpha)}=(0,-\alpha,1)$ and
$\ve{c}_{\xi(\alpha)}=(1,0,0)$.  Special cases are: $\xi(0)=\xi$ and
$\xi(1)=\xi'$. Phase~2 on the right side is the $\xi'_{2}$-approximant.

The tiling for this phasonic phase boundary from $\xi(0.8)$ to $\xi'_{2}$ is
calculated as explained above and shown in figure~\ref{fig:phaseboundary2}(b).
For the transition function we used:
\begin{equation}
f(x)=\left\{\begin{array}{l@{\;,\;}l}
0 & x\leq 0\\
\frac{\arctan(x)}{\pi}+\frac{1}{2} & x>0.
\end{array}\right.
\end{equation}
In the region of the $\xi(0.8)$-phase ($x<0$) the cut plane is flat, since in
the $\xi(\alpha)$-phases the phasonic degree of freedom is not excitable, just
as in the $\xi$- or the $\xi'$-phase. In the region of the $\xi'_{2}$-phase
($x>0$) excitations of the phasonic degree of freedom show as bendings of the
phason-planes.

As a direct consequence of the hyperspace description the orientation of the
phasonic phase boundary is completely determined by the involved approximant
phases. On the other side, by observing this orientation we can identify the
approximant phases, labelled by the parameter $\alpha$.

\section{Dislocations: Energetic consideration}

In quasicrystals and approximants dislocations do exist and are characterised
by a unique Burgers vector $\ve{b}$, which now is a vector of the
hyperlattice. A dislocation is accompanied by a phonon- and a phason-strain field,
due to the phononic part $\ve{b}_{\parallel}=\pi_{\parallel}\ve{b}$ and
the phasonic part $\ve{b}_{\perp}=\pi_{\perp}\ve{b}$ ($\pi_{\perp}$ projects
on the orthogonal complement of $E$), respectively, of the Burgers vector. If
the phasonic part vanishes, the dislocation is a classical one and can be
described without the hyperspace methods. Since the lattice constants of
periodic approximants are large, classical dislocations have gigantic Burgers
vectors with huge phonon-strains. They are energetically unfavourable and will
not be observed.

By extending the linear theory of elasticity to hyperspace (approximating the
phasonic degree as continuous), the line energy $E$ of a dislocation grows
quadratically with increasing lengths $\|\ve{b}_{\parallel}\|$ and
$\|\ve{b}_{\perp}\|$.\footnote{There is an ongoing discussion whether the
  elastic energy grows linear (locked state) or quadratically (unlocked state)
  in the phasonic strain, see the review by \cite{itapdb:Edagawa2001}. However
  at higher temperatures, when dislocations can form and move, the unlocked
  state is entropically stabilised.}
Assuming isotropy in the phononic and phasonic part, $E$ can be expressed as:
\begin{equation}
\label{eq:energy}
E=c_{\text{phon}}\|\ve{b}_{\parallel}\|^{2}+
c_{\text{phas}}\|\ve{b}_{\perp}\|^{2}+
c_{\text{coupl}}\|\ve{b}_{\parallel}\|\|\ve{b}_{\perp}\|.
\end{equation}
Besides a phononic contribution with elastic constant $c_{\text{phon}}$ and a
phasonic contribution with $c_{\text{phas}}$, a coupling term is present with
$c_{\text{coupl}}$.\footnote{In several papers the coupling term is neglected,
  although its appearance is a consequence of the linear theory of
  elasticity.}

For estimating the different contributions in (\ref{eq:energy}) the elastic
constants in the approximant phases are assumed to be comparable to those of
i-Al-Pd-Mn. For the icosahedral phase the phononic elastic constants have been
determined by \cite{itapdb:Amazit1995} from
sound-propagation. The phasonic elastic constants have been measured with
neutron and x-ray scattering experiments by
\cite{itapdb:Letoublon2001}. From the experimental values it can be
concluded, that $c_{\text{phas}}$ is a few orders of magnitudes smaller than
$c_{\text{phon}}$. Furthermore computer simulations
\cite[]{itapdb:Koschella2002} suggest that $c_{\text{coupl}}$ has about
the same value as $c_{\text{phas}}$. So the largest contribution to $E$ comes
from the phononic part. In the tiling picture this means that it is
energetically more favourable to rearrange parts of the tiling than to deform
tiles.

\section{Dislocations in the $\xi'_{n}$-phases}

The construction method for a dislocation in hyperspace has been discussed
e.g.\ by \cite{itapdb:Bohsung1989b}. A dislocation
with Burgers vector $\ve{b}$ forces upon the cut plane a displacement field
$\ve{u}$ with a phononic and a phasonic part. In a first approximation the
displacement field can be distributed isotropically in physical space and
becomes in spherical coordinates:
\begin{equation}
\ve{u}(r,\varphi)=\frac{\varphi}{2\pi}\ve{b}.
\end{equation}

\subsection{Metadislocations in the three-dimensional model}
\begin{figure}\begin{center}
\includegraphics[width=5cm]{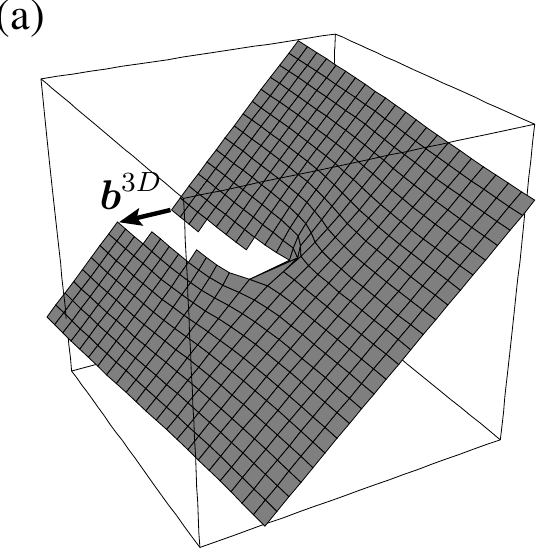}\quad
\includegraphics[width=5cm]{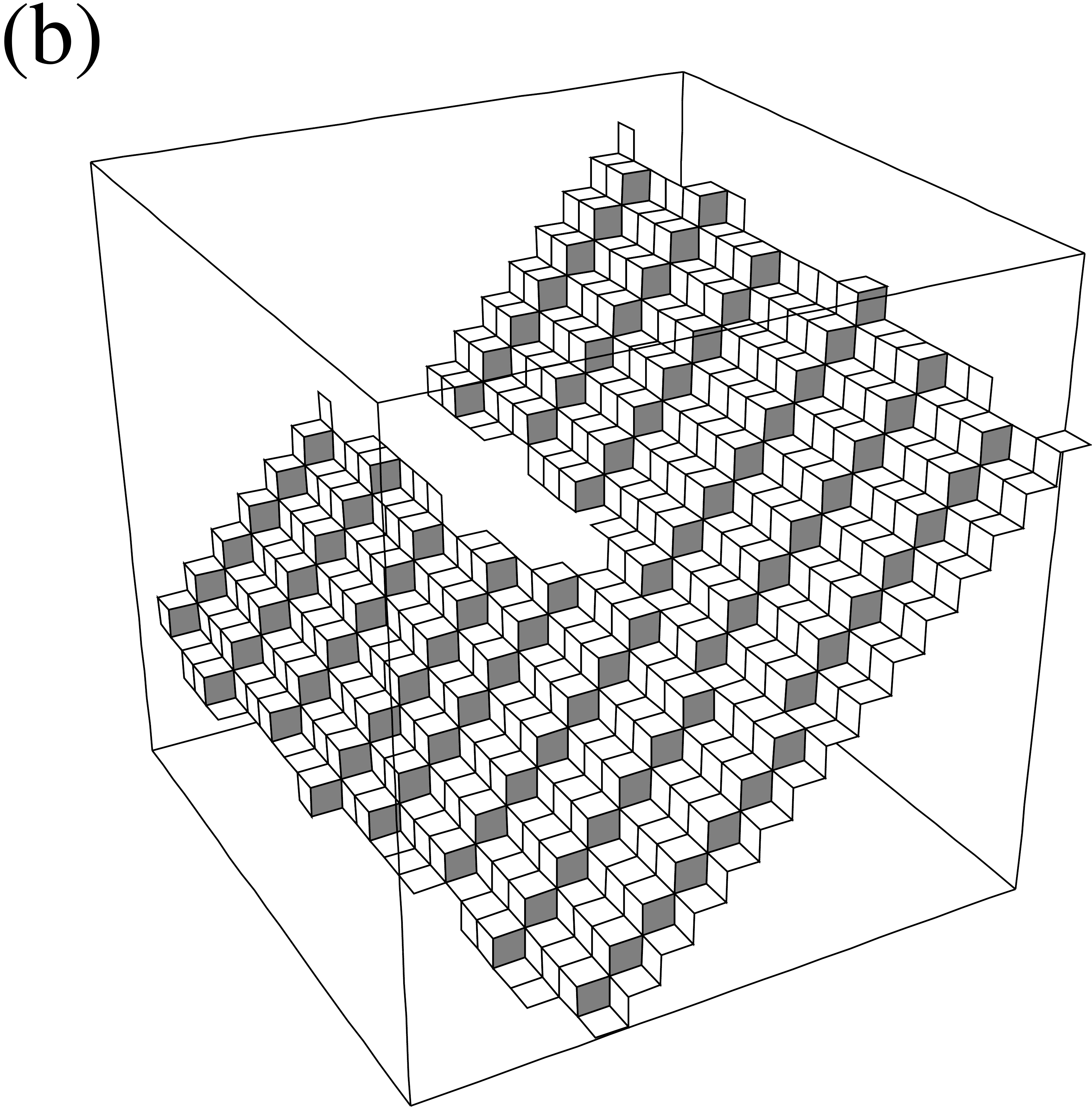}\quad
\includegraphics[width=5cm]{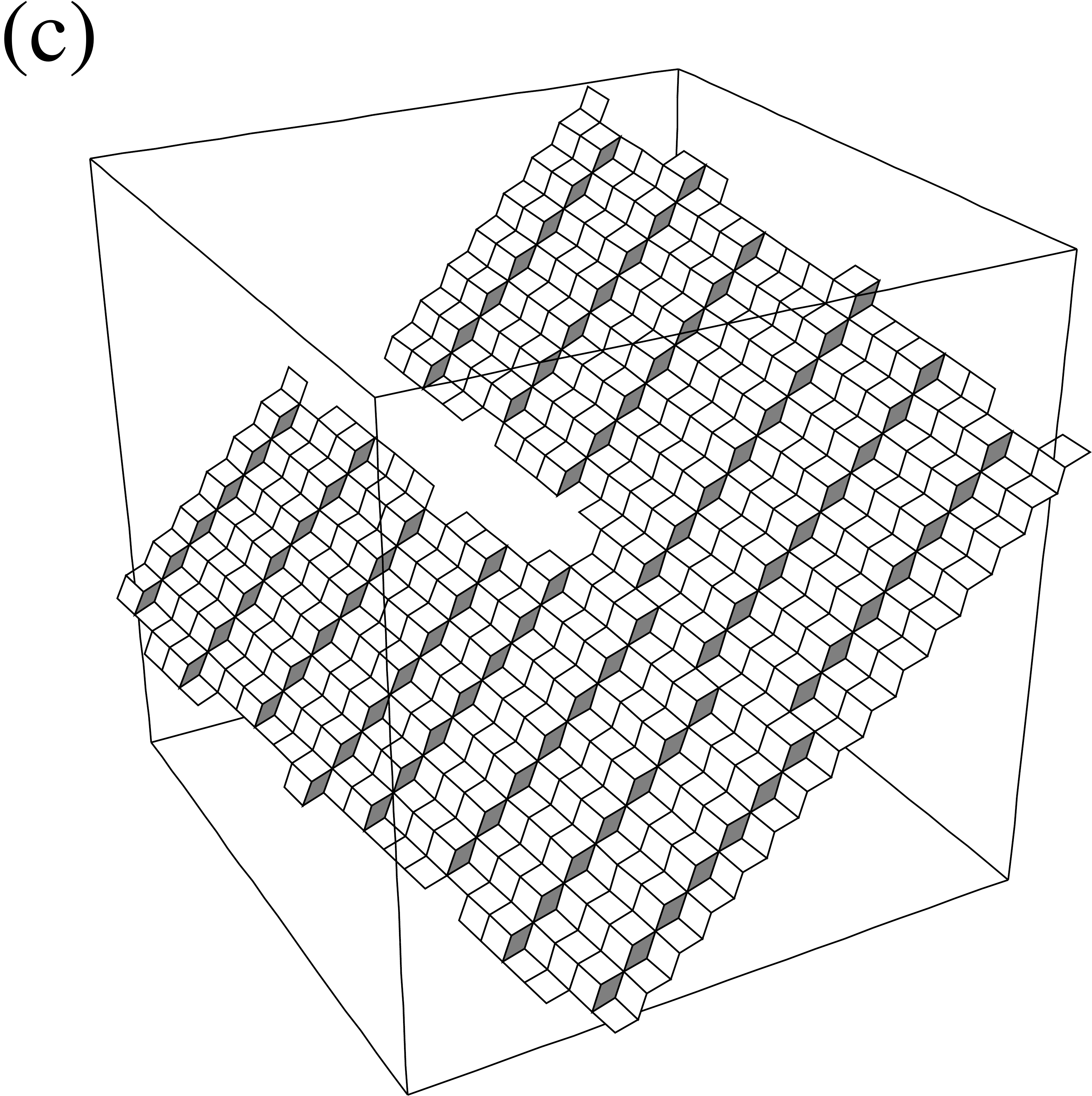}\\[0.5cm]
\includegraphics[width=4.5cm]{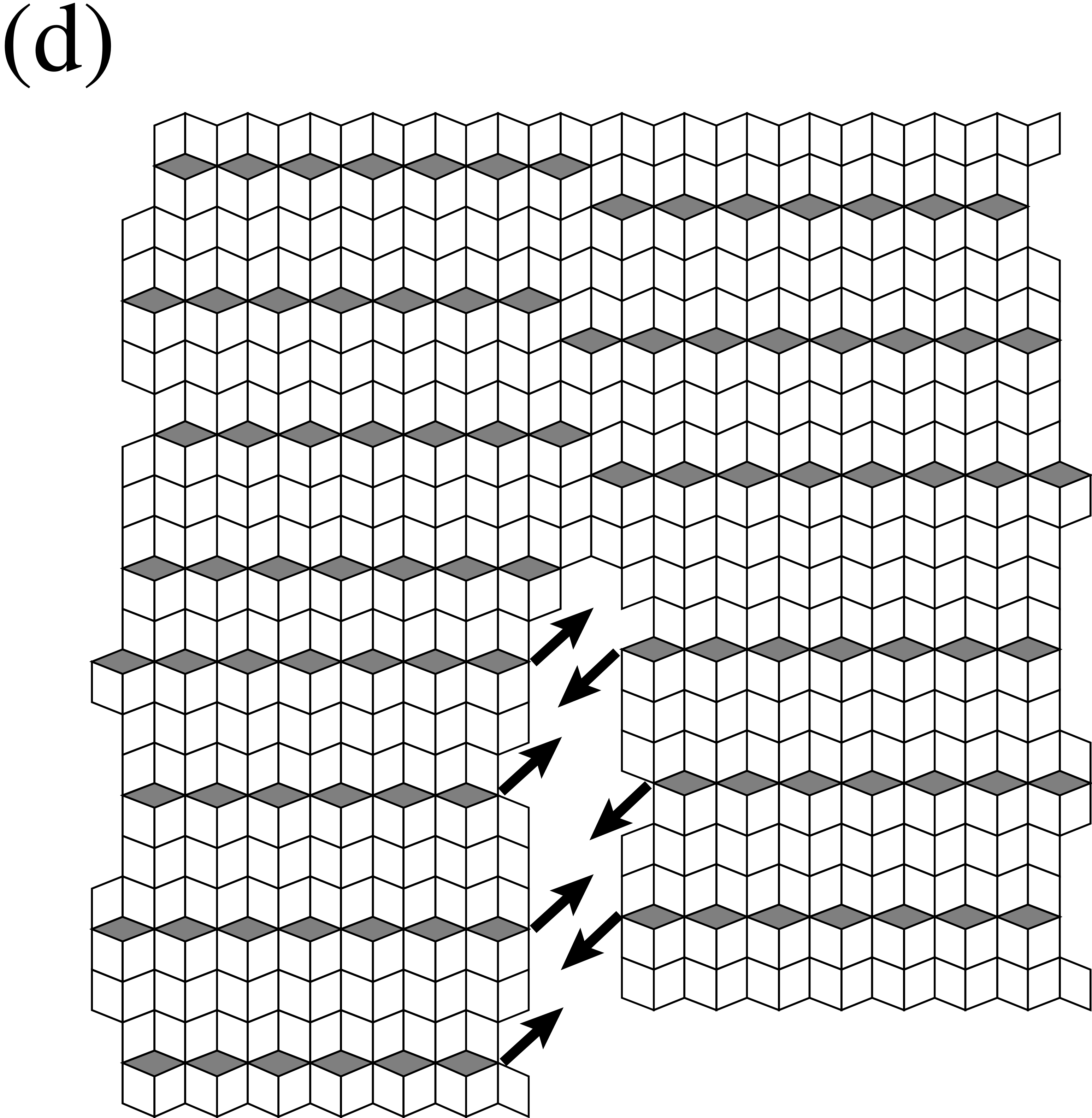}\quad
\includegraphics[width=4.5cm]{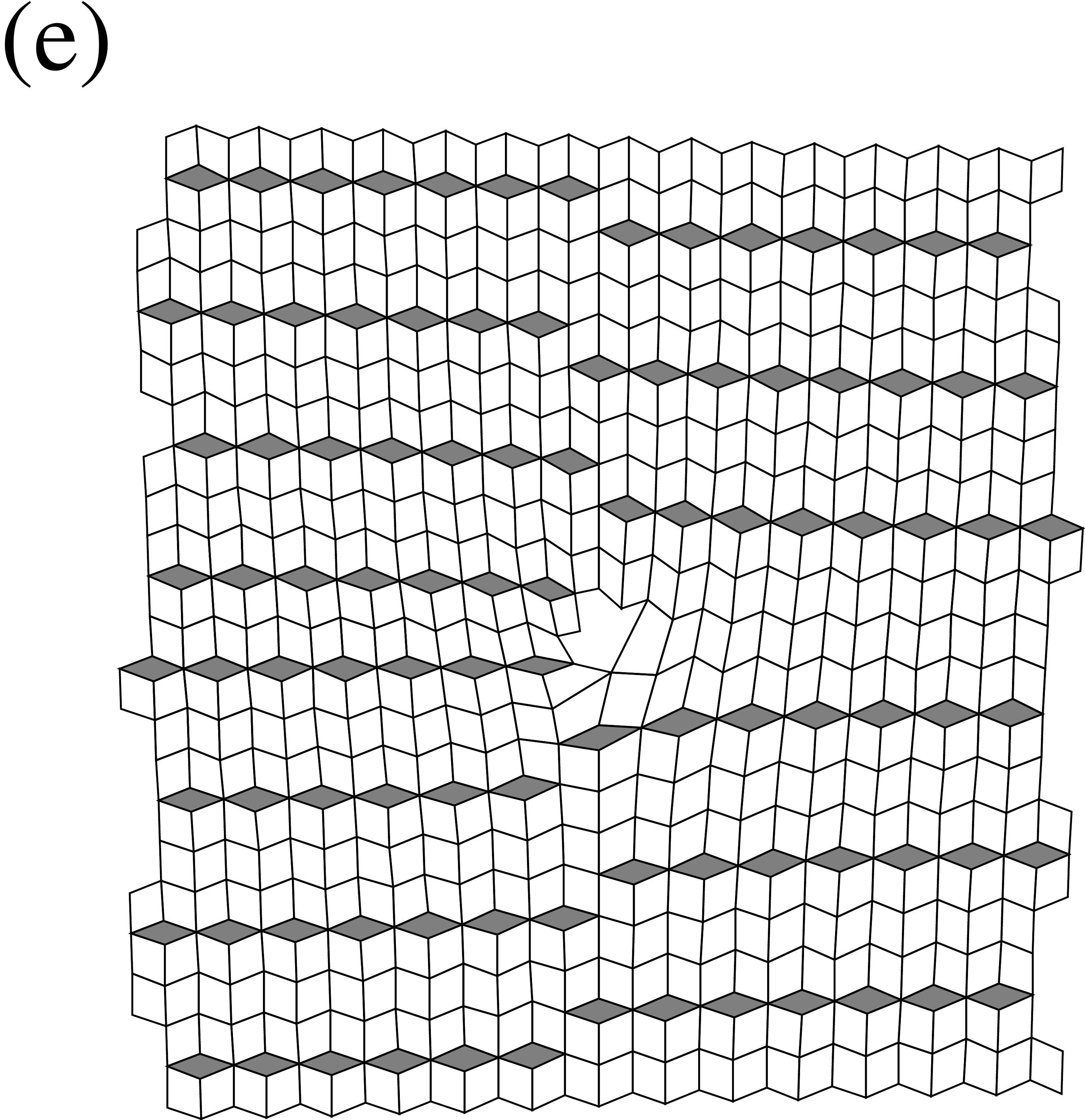}\quad
\includegraphics[width=6cm]{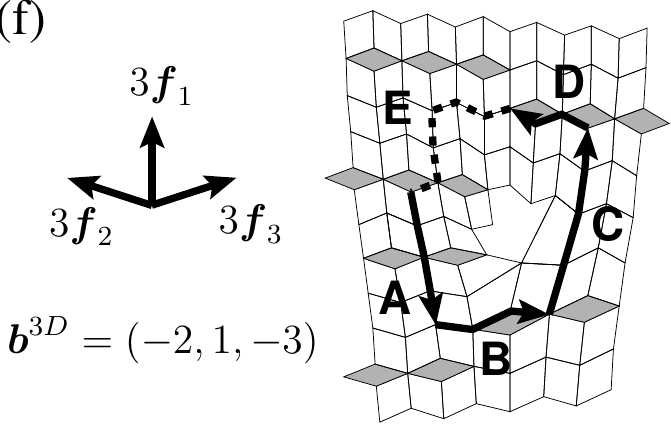}
  \caption{Construction of a dislocation in the three-dimensional
    model. (a)~The displaced cut plane $\mathcal{E}$ of the $\xi'_{3}$-phase
  around the dislocation core. (b)~Staircase surface of those atomic volumes,
  that are cut by $\mathcal{E}$. (c)~Projection of the staircase surface onto
  the projection plane $E$. (d)~The new tiling around the dislocation core
  with rearrangements of the tiles due to phason-strain. (e)~The gap in the
  tiling is closed by phonon-strain. (f)~The Burgers vector $\ve{b}^{3d}$ can
  be obtained by performing a Burgers circuit. \label{fig:dislocation}}
\end{center}\end{figure}

The steps involved in the construction of a dislocation are visualised in
figure~\ref{fig:dislocation}. The distorted cut plane has a strain jump of
$\ve{b}^{3D}$ along a radial line as shown in
figure~\ref{fig:dislocation}(a). The middle points of the atomic volumes
selected by the cut plane form a staircase surface as shown in
figure~\ref{fig:dislocation}(b). This surface is projected with
$\pi^{3D}_{\parallel}$ onto the projection plane
(figure~\ref{fig:dislocation}(c)). Up to now only the phason-strain of the
dislocation has been taken regard of and results in a rearrangement of the
tiles. The gap in the tiling is caused by the jump line in the cut plane. In a
last step the phonon-strain is introduced (figures~\ref{fig:dislocation}(d),
(e)), closing the gap and deforming the tiles. In
(figure~\ref{fig:dislocation}(f)) a Burgers circuit is constructed. The sides
of the tiles correspond to projected basis vectors $\ve{f}_{i}$ of the
hyperlattice. In the example the lines A, C and B, D cancel each other, only E
is left and is lifted to the Burgers vector
$\ve{b}^{3D}=(-2,1,-3)=-2\ve{e}^{3D}_{1}+\ve{e}^{3D}_{2}-3\ve{e}^{3D}_{3}$.

Consider now an arbitrary dislocation with $\ve{b}^{3D}=(b_{1},b_{2},b_{3})$,
$b_{i}\in\mathbbm{Z}$. Its phononic component
\begin{equation}
\ve{b}_{\parallel}=\pi^{3D}_{\parallel}\ve{b}^{3D}=
\frac{1}{5}\sqrt{10}\left(\begin{array}{c}
\frac{1}{2}\sqrt{\tau+2}(b_{2}-b_{3})\\
b_{1}+\frac{1}{2}\tau^{-1}(b_{2}+b_{3})\\
\end{array}\right).
\end{equation}
must be small for low dislocation energy, as shown in the last section. This
is achieved, if $b_{2}=b_{3}$ and $\frac{b_{2}}{b_{1}}\approx -\tau$. The best
approximating rational values for $\tau$ are given by fractions
$\frac{F_{m}}{F_{m-1}}$ of successive Fibonacci numbers
$(F_{m})_{m\in\mathbbm{N}}=(1,1,2,3,5,8,13,21,\ldots)$ defined by
$F_{m+1}=F_{m-1}+F_{m}$ with start values $F_{1}=F_{2}=1$. They have the
following properties, which can be proved by induction:
\begin{subequations}
\label{eq:properties_fm}
\begin{eqnarray}
&\text{(i)}  \quad& F_{m}\tau+F_{m-1}=\tau^{m}\\
&\text{(ii)} \quad& F_{m+1}-F_{m}\tau=(-\tau)^{-m}\\
&\text{(iii)}\quad& F_{m}=\frac{1}{5}\sqrt{5}(\tau^{m}-(-\tau)^{-m})
\end{eqnarray}
\end{subequations}

Therefore the best candidates for Burgers vectors of low energy dislocations are
\begin{equation}
\label{eq:burgersvector}
\ve{b}^{3D}=\left(\begin{array}{c}
F_{m-1}\\
-F_{m}\\
-F_{m}\\
\end{array}\right)\quad\text{with}\quad
\ve{b}_{\parallel}=\frac{1}{5}\sqrt{10}\left(\begin{array}{c}
0\\F_{m-1}-\tau^{-1}F_{m}
\end{array}\right)
\end{equation}
and length:
\begin{equation}
\|\ve{b}_{\parallel}\|=\tau^{-m}\frac{1}{5}\sqrt{10}\,l^{3D}=\tau^{-m}\cdot
1.26\text{ nm}.
\end{equation}
With (\ref{eq:transition5D6D}) and (\ref{eq:transition3D5D}) we get the 6D
Burgers vector:
\begin{equation}
\ve{b}^{6D}=T_{5D}^{6D}T_{3D}^{5D}\ve{b}^{3D}=
(0,0,-F_{m-2},F_{m-1},F_{m-2},F_{m-1}).
\end{equation}
The phasonic component for the six-dimensional hyperspace can be calculated
with (\ref{eq:properties_fm}):
\begin{eqnarray}
\|\ve{b}_{\perp}\|&=&\sqrt{\|\ve{b}^{6D}\|^{2}-\|\ve{b}_{\parallel}\|^{2}}=
\sqrt{2\left[(F_{m-1})^{2}+(F_{m-2})^{2}\right](l^{6D})^{2}-
\frac{2}{5}\tau^{-2m}(l^{3D})^{2}}\notag\\
&=&\tau^{m-3}\frac{1}{5}\sqrt{10}\,l^{3D}=\tau^{m-3}\cdot
1.26\text{ nm},
\end{eqnarray}
yielding the strain accommodation parameter $\zeta$, i.e.\ the ratio of the
phasonic and the phononic component of the Burgers vector
\cite[]{itapdb:Feuerbacher1997b}:
\begin{equation}
\zeta=\frac{\|\ve{b}_{\perp}\|}{\|\ve{b}_{\parallel}\|}=\tau^{2m-3}.
\end{equation}

\begin{figure}\begin{center}
\includegraphics[height=6cm]{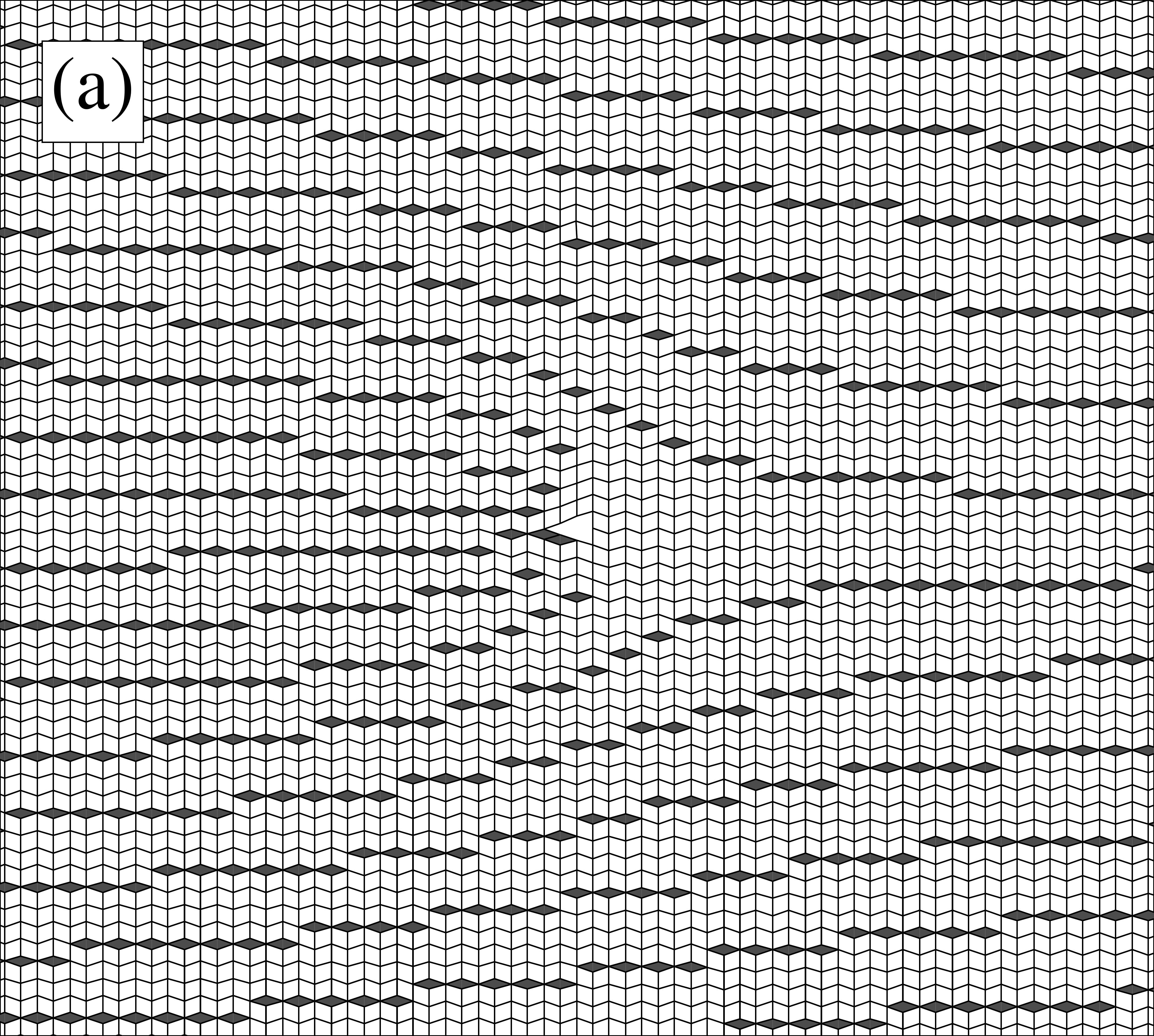}\qquad
\includegraphics[height=6cm]{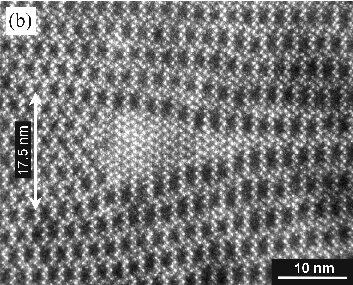}\\[0.5cm]
\includegraphics[height=6cm]{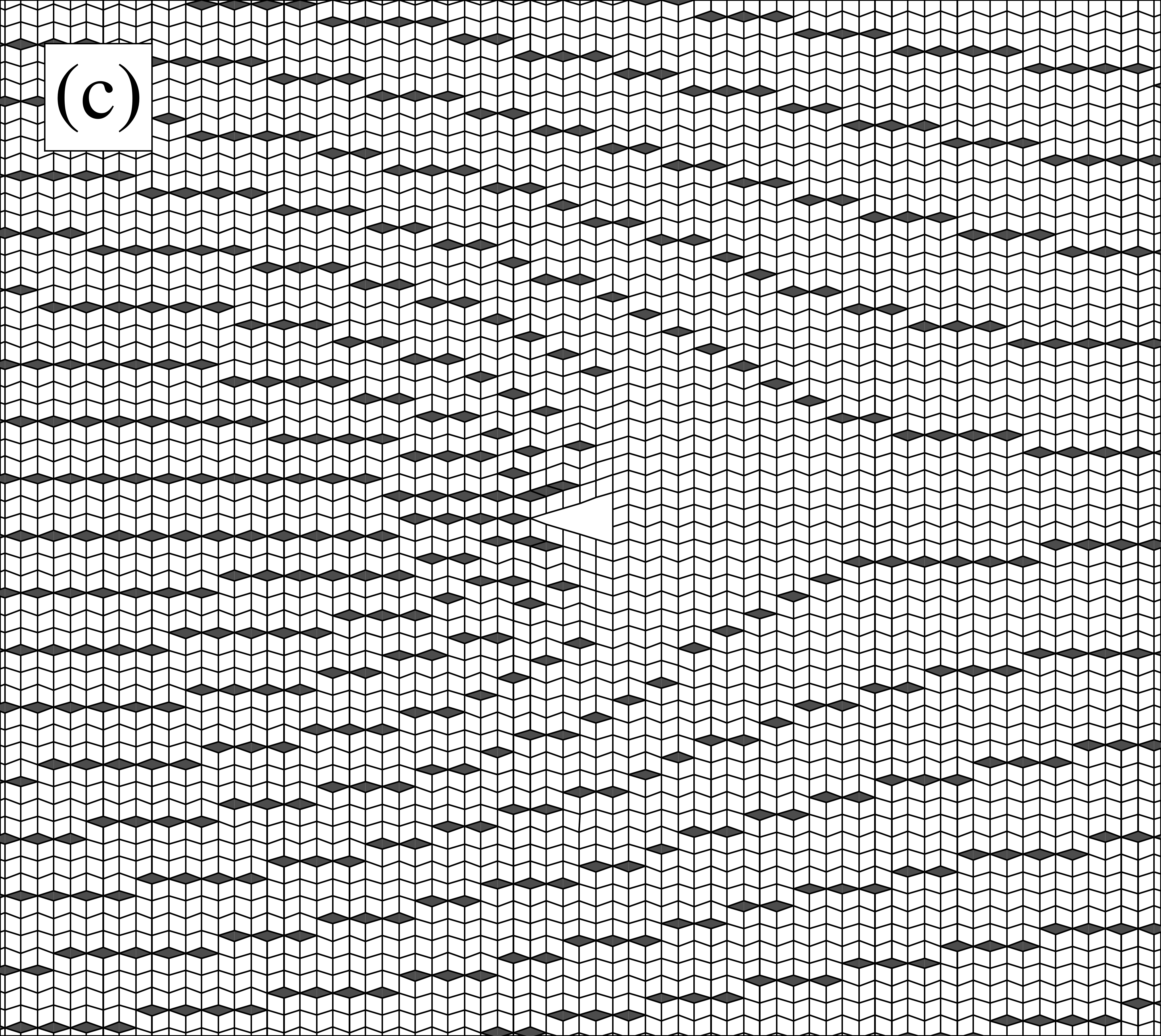}\qquad
\includegraphics[height=6cm]{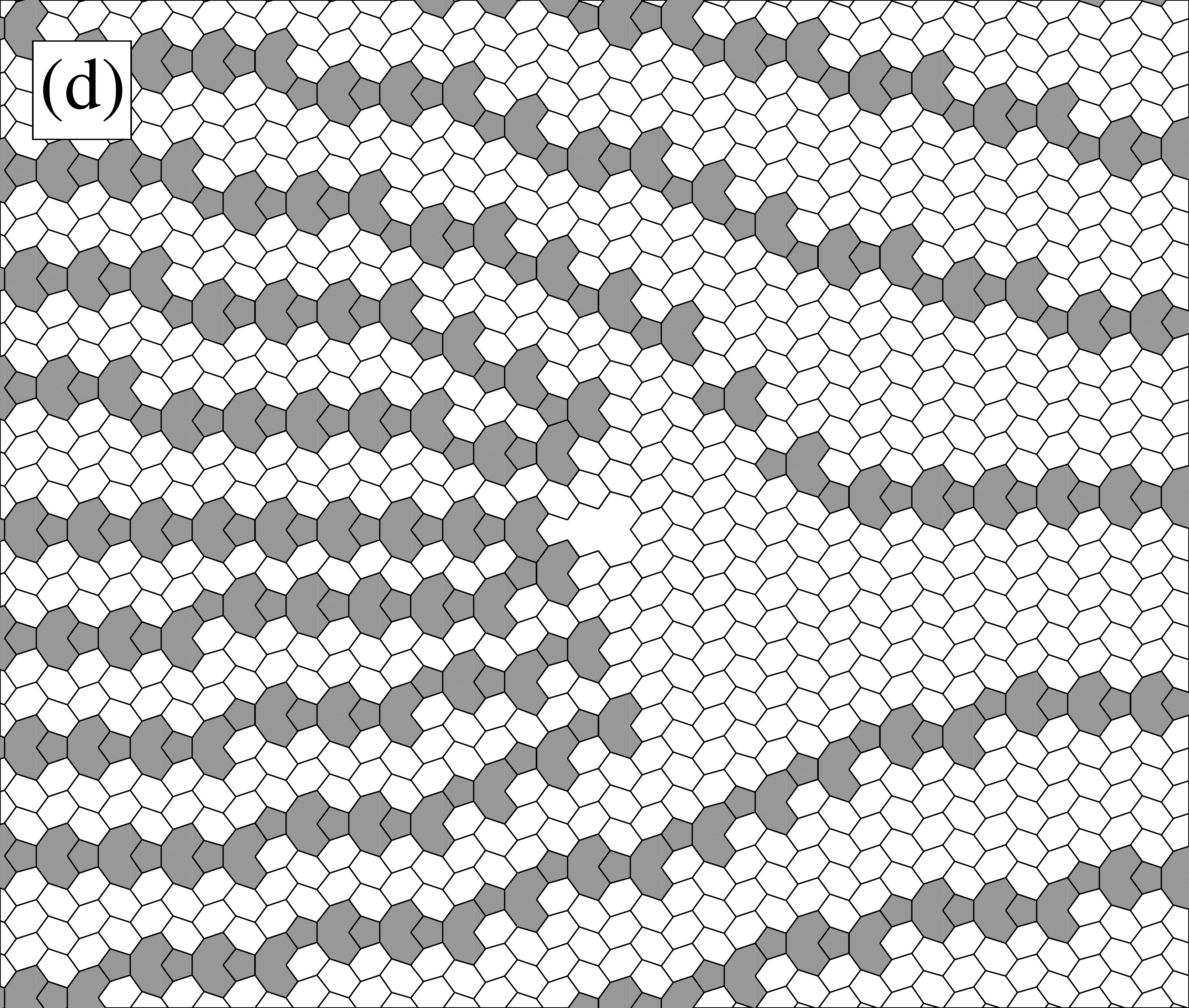}
  \caption{(a)~Tiling of a type 4 metadislocation in the
    $\xi'_{4}$-phase. (b)~Transmission electron micrograph of a type 4
  metadislocation in the $\xi'_{2}$-phase, courtesy by H. Klein and
  M. Feuerbacher. (c)~Tiling of a type 5 metadislocation in the
  $\xi'_{4}$-phase. (d)~For comparison the calculated hexagon/pentagon/nonagon
  tiling of a type 4 metadislocation in the $\xi'_{4}$-phase is
  shown. \label{fig:metadislocation}}
\end{center}\end{figure}

Tilings for dislocations with $m=4$ and $m=5$ are shown in
figures~\ref{fig:metadislocation}(a) and (c). The dislocation cores appear as
triangles which can be used for Burgers circuits.  Phason-planes come in from
the left side of the tiling, ending at the flat side of the triangle, their
number being equal to that of the tiles there, which is $2F_{m}$. Whereas the
phononic part of the Burgers vector is much smaller than the side length
$l^{3D}$ of the rhombs, the phasonic part is so large that it leads to a
bending of the phason-planes over a huge area of the tiling.

The arrangement of the phason-lines can be viewed as a metastructure which, in
the simple case of a dislocation free $\xi'_{n}$-phase is a striped centred
rectangular lattice. A partial dislocation in the tiling leads to a dislocation
in the striped metastructure, named metadislocation. A dislocation with a
three-dimensional Burgers vector as in (\ref{eq:burgersvector}) is called
metadislocation of type $m$. In contrary to the dislocation constructed in
hyperspace, a metadislocation is an ordinary dislocation in the metastructure
with a two-dimensional Burgers vector.

The Burgers vector $\ve{b}_{\text{meta}}$ of a type $m$ metadislocation is
oriented in the direction of the lattice vector
$\ve{c}_{\xi'_{n}}=\pi^{3D}_{\parallel}\ve{c}^{3D}_{\xi'_{n}}$. Since a
lattice cell contains two phason-planes, a type $m$ metadislocation has
$F_{m}$ inserted rows of lattice cells in this direction, leading to a Burgers
vector
\begin{equation}
\ve{b}_{\text{meta}}=F_{m}\pi^{3D}_{\parallel}\ve{c}^{3D}_{\xi'_{n}}=
\frac{F_{m}}{5}\sqrt{10}\left(\begin{array}{c}
0\\ 2n+\tau^{-1}\\
\end{array}\right)
\end{equation}
of length
\begin{equation}
\|\ve{b}_{\text{meta}}\|=l^{3D}F_{m}(2n+\tau^{-1})=F_{m}(2n+\tau^{-1})\cdot
1.26\text{ nm}.
\end{equation}

The most frequent dislocations are the type 4 metadislocations in the
$\xi'_{2}$-phase. They have been observed by electron microscopy
\cite[]{itapdb:Klein1999}, see figure~\ref{fig:metadislocation}(b).
From our theory we derive the lengths $\|\ve{b}_{\parallel}\|=0.184$~nm and
$\|\ve{b}_{\perp}\|=2.04$~nm ($\zeta=11.1$) for the dislocation and
$\|\ve{b}_{\text{meta}}\|=17.5$~nm for the metadislocation, which is larger by
two orders of magnitude. For comparison we present the metadislocation also in
the hexagon/pentagon/nonagon-tiling (figure~\ref{fig:metadislocation}(d)).
Transmission electron micrographs of type 3, 5 and 6 metadislocations have been
published, too, by \cite{itapdb:Klein2003}.

\subsection{Metadislocations in the five-dimensional model}

\begin{figure}\begin{center}
\includegraphics[height=5cm]{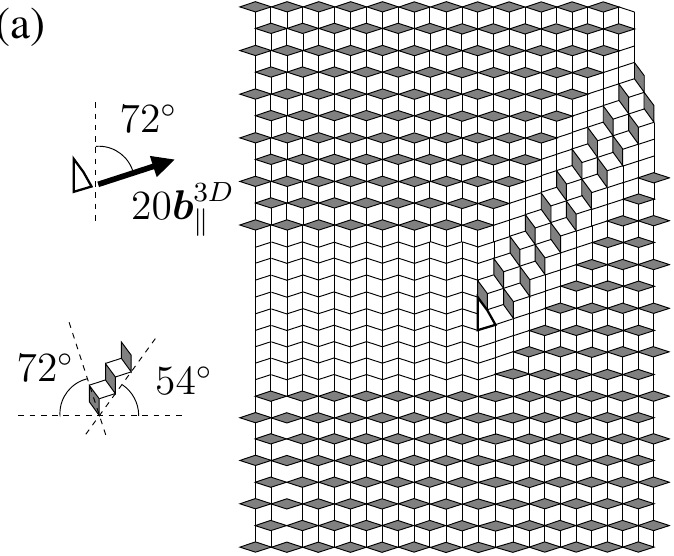}\qquad
\includegraphics[height=5cm]{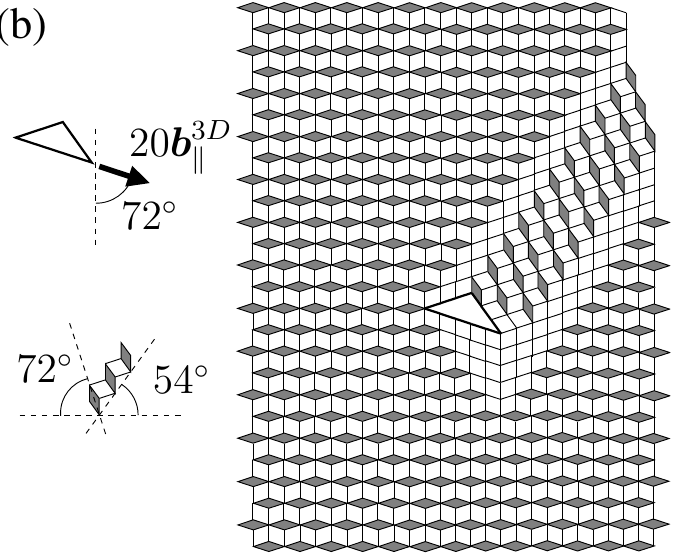}\\[0.5cm]
\includegraphics[height=6cm]{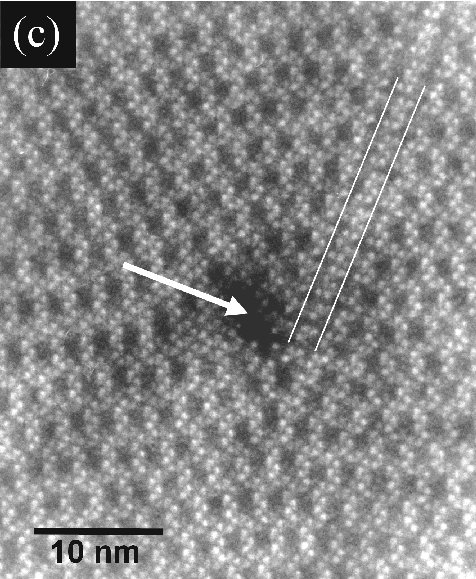}\qquad
\includegraphics[height=6cm]{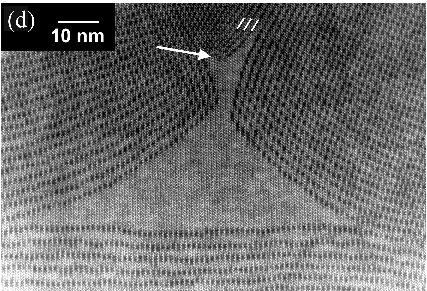}
  \caption{Tilings of metadislocations in the $\xi'_{1}$-phase with Burgers
    vectors (a)~$\ve{b}^{5D}=(2,-1,2,0,0)$ and
  (b)~$\ve{b}^{5D}=(0,3,3,0,5)$. The phason-strain field is not distributed
  isotropic. The phononic component of the Burgers vector is drawn magnified
  20 times, since it is too small otherwise. (c)~and (d) show transmission
  electron micrographs of these two metadislocations, courtesy by H. Klein and
  M. Feuerbacher. \label{fig:metadislocation2}}
\end{center}\end{figure}

Other kinds of metadislocations reported by
\cite{itapdb:Klein2003} also show a large scale arrangement of
phason-lines, but require the five-dimensional hyperspace for description,
since phason-lines occur in more than one orientation. Here we will discuss
two different types of metadislocations:
\begin{enumerate}
\item Metadislocations with Burgers vector of the form:
  $\ve{b}^{5D}=(F_{m},-F_{m-1},F_{m},0,0)$: The phononic component is:
  \begin{equation}
  \ve{b}_{\parallel}=\pi^{5D}_{\parallel}\ve{b}^{5D}=
  \frac{1}{5}\sqrt{10}(-\tau)^{-m}\left(\begin{array}{c}
  \cos(72^{\circ})\\
  \sin(72^{\circ})
  \end{array}\right).
  \end{equation}
  with length $\|\ve{b}_{\parallel}\|=\tau^{-m}\cdot1.26$~nm. The Burgers
  vector is rotated by $72^{\circ}$ compared to the metadislocation in
  (\ref{eq:burgersvector}). A tiling of such a metadislocation with $m=3$ is
  shown in figure~\ref{fig:metadislocation2}(a). Phason-lines now appear in
  two different orientations. $F_{m}$ rows of tilted phason-lines ending at
  the dislocation core cannot be explained within the three-dimensional
  model.
\item Metadislocations with Burgers vector of the form:
  $\ve{b}^{5D}=(0,F_{m-1},F_{m-1},0,F_{m})$: The phononic component is:
  \begin{equation}
  \ve{b}_{\parallel}=
  \frac{1}{5}\sqrt{10}(-\tau)^{-m-1}\left(\begin{array}{c}
  \cos(-72^{\circ})\\
  \sin(-72^{\circ})
  \end{array}\right).
  \end{equation}
  with length $\|\ve{b}_{\parallel}\|=\tau^{-m-1}\cdot1.26$~nm. A tiling with
  $m=5$ is shown in figure~\ref{fig:metadislocation2}(b). This time $F_{m-1}$
  rows of tilted phason-lines end at the dislocation core.
\end{enumerate}

For transmission electron micrographs of the two discussed metadislocations see
figure~\ref{fig:metadislocation2}(c) and (d). In the first figure the gap
extending to the left in the tiling has been closed by the motion of the
phason-planes. In the second figure the metadislocation seems to have moved
up, while building a sack-shaped area of $\xi'$-phase below. On the
sides of this area, five phason-planes are cut into halves by a region of
$\xi'$-phase.

\section{Metadislocations in the $\xi'$-phase}

\begin{figure}\begin{center}
\includegraphics[height=5cm]{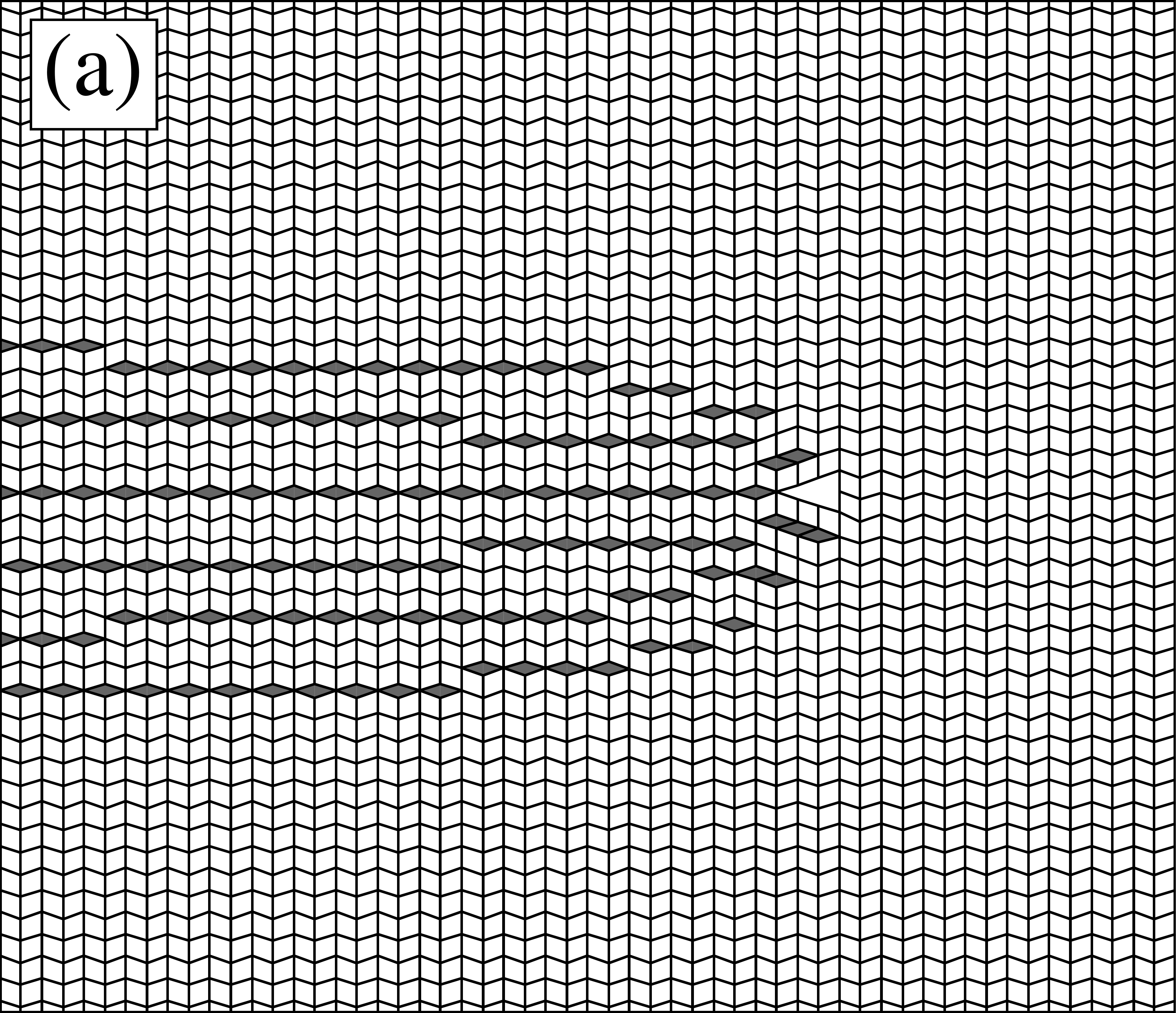}\qquad
\includegraphics[height=5cm]{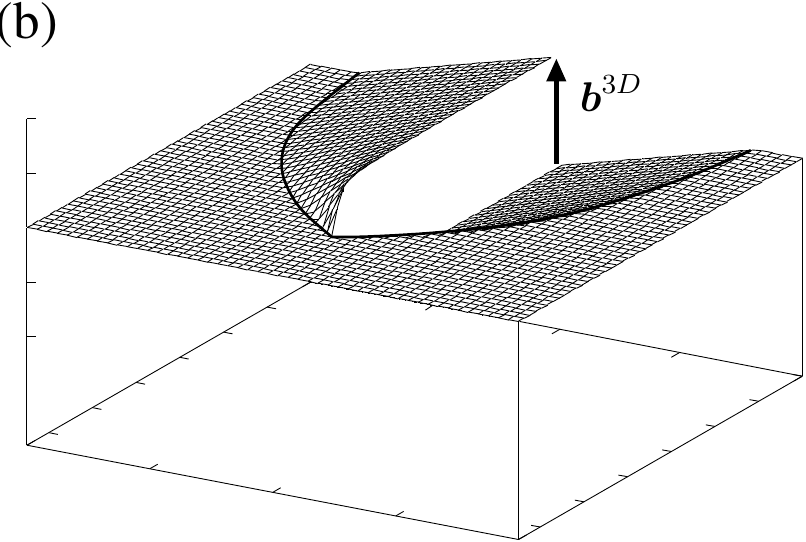}
  \caption{(a)~A tiling of a metadislocation in the $\xi'$-phase with six
    inserted phason-planes. In the dislocation tail a strip of
  $\xi'_{3}$-phase is created. (b)~Cut plane for the metadislocation in the
  $\xi'$-phase. The phason-strain field is not relaxed isotropic, but
  concentrated in the dislocation tail, where the phason-lines are
  observed. \label{fig:metadislocation3}}
\end{center}\end{figure}

Although phasonic degrees of freedom are not excitable {\em continuously} in
the $\xi'$-phase, metadislocations with phasonic components, which cause a
{\em noncontinuous} local phason-strain do exist and have been observed
experimentally by \cite{itapdb:Klein2003}. They
can be described in the three-dimensional hyperspace. To minimise the
dislocation energy, their Burgers vectors $\ve{b}^{3D}$ are the same as in the
$\xi'_{n}$-phases (\ref{eq:burgersvector}). For a tiling see
figure~\ref{fig:metadislocation3}(a) in the case $m=4$. Similar to
metadislocations in the $\xi'_{n}$-phases, a tail of six inserted
phason-planes ends at the flat sides of the dislocation core. Towards the
dislocation core the phason-planes converge, while far away they run
parallel. Contrary to metadislocations in the $\xi'_{n}$-phases the
phason-strain now is not distributed isotropically around the dislocation
core, but concentrated in the region, where phason-lines are observed.

The shape of the cut plane for the metadislocation in the three-dimensional
model is shown in figure~\ref{fig:metadislocation3}(b), where the coordinate
system is rotated, so that the cut plane $\mathcal{E}_{\xi'}$ of the
$\xi'$-phase lies horizontally. Because of its small phononic component, the
Burgers vector is almost perpendicular to $\mathcal{E}_{\xi'}$.

We place the dislocation in the origin and the strain jump of $\ve{b}^{3D}$
along the $y$-axis, along a line in the middle of the strip of phason
lines. Inside the dislocation tail and far away from the core the strain field is
\begin{equation}
\ve{u}^{3D}(x,y)=\frac{x}{2}\left(\frac{1}{|x|}-\frac{1}{\lambda}\right)
\ve{b}^{3D}.
\end{equation}

The parameter $\lambda$ controls the width of the dislocation tail. To
simulate a convergence of the dislocation tail $\lambda=\lambda(y)$ must be
designed decreasing with distance from the dislocation core. Outside of the
dislocation tail, the cut plane is parallel to $\mathcal{E}_{\xi'}$. The
region with the phason lines can be viewed as a very thin strip of
newly created $\xi'_{n}$-phase, embedded in surrounding $\xi'$-phase. We
suggest, that these dislocations move in the direction of their tail, because
for the movement perpendicular to it the entire tail would have to be dragged
along. The movement is pure climb, because it is perpendicular to the burgers
vector and the dislocation line. There are indications
\cite[]{itapdb:Mompiou2004}, that in the i-Al-Pd-Mn quasicrystal dislocations 
can move by pure climb. Furthermore the most often observed dislocations in
i-Al-Pd-Mn have the same burgers vectors as the metadislocations in the
$\Xi$-phases \cite[]{itapdb:Rosenfeld1995}.

Finally it should be mentioned that \cite{itapdb:Feuerbacher2004} also
observed other dislocations in the $\xi'$-phase with Burgers vectors that do
not lie perpendicular to the vector $\ve{b}^{6D}_{\xi'}$. These cannot be
pictured as two-dimensional tilings, but require a true three-dimensional
description in physical space.

\section{Conclusion}

In this paper we have introduced a simple projection formalism in
five-dimensional hyperspace for the description of the i-(Al-Pd-Mn)
approximants $\xi$, $\xi'$ and $\xi'_{n}$ as two-dimensional tilings. The
tilings correspond to the projections of clusters columns. In most cases the
formalism can even be restricted to a three-dimensional hyperspace. The
tilings are generated by cuts and projections through a hyperspace which is
partitioned into atomic hypervolumes.

We have shown that phasonic degrees of freedom in form of continuous
displacements of the cut space do exist in these phases and can be either
excitable or not. They play a fundamental role for phasonic phase boundaries
as well as dislocations. In the case of the $\xi$- and $\xi'$-phase no
phasonic degrees of freedom are excitable, while in the $\xi'_{n}$-phases
there is exactly one excitable phasonic degree of freedom, which is connected
to the bending of the phason-planes.

Nevertheless metadislocations in the $\xi'$-phase can exist, creating
phason-planes and forming a thin strip of $\xi'_{n}$-phase in their tails
while moving by climb. In this way, a consecutive motion of metadislocations
through the $\xi'$-phase induces a phase transformation to a $\xi'_{n}$-phase,
making the phasonic degree of freedom excitable. The same metadislocations
continue to exist in the $\xi'_{n}$-phases.

It is possible that phasonic degrees of freedom and dislocations with phasonic
components also occur in various other complex metallic alloys, that have
connections to quasiperiodic phases. Thus restrictions for ordinary
dislocations, that are imposed by the large unit cells, can be overcome.

\addcontentsline{toc}{chapter}{References}
\bibliographystyle{abbrvnat}

{\small
\begin{thebibliography}{20}

\bibitem[{Amazit} et~al.(1995)]{itapdb:Amazit1995}
{\sc {Amazit},~Y., {Fischer},~M., {Perrin},~B.,} and {\sc {Zarembowitch},~A.},
\newblock 1995, {\it {P}roccedings of the 5th {I}nternational {C}onference on
{Q}uasicrystals}, edited by C.~{Janot} and R.~{Mossieri} (Singapore: {W}orld
{S}cientific), p.584.

\bibitem[{Beraha} et~al.(1997)]{itapdb:Beraha1997}
{\sc {Beraha},~L., {Duneau},~M., {Klein},~H.,} and {\sc {Audier},~M.},
\newblock 1997, {\it Phil.\ Mag.}~A, {\bf 76}, 587.

\bibitem[{Bohsung} and {Trebin}(1989)]{itapdb:Bohsung1989b}
{\sc {Bohsung},~J.,} and {\sc {Trebin},~H.-R.},
\newblock 1989, {\it Introduction to the Mathematics of Quasicrystals}, edited
by M.V.~{Jari\'c} (London: Academic Press), p.183.

\bibitem[{Boudard} et al.(1996)]{itapdb:Boudard1996b}
{\sc {Boudard},~M., {Klein},~H., {de~Boissieu},~M., {Audier},~M.,} and
{\sc {Vincent},~H.},
\newblock 1996, {\it Phil.\ Mag.}~A, {\bf 74}, 939.

\bibitem[{Duneau} and {Audier}(1994)]{itapdb:Duneau1994}
{\sc {Duneau},~M.,} and {\sc {Audier},~M.},
\newblock 1994, {\it Lectures on quasicrystals}, edited by F.~{Hippert} and
D.~{Gratias} (Les Ulis: Les Editions de Physique), p. 283.

\bibitem[{Edagawa}(2001)]{itapdb:Edagawa2001}
{\sc {Edagawa},~K.},
\newblock 2001, {\it Mater.\ Sci.\ Eng.}, {\bf A309}, 528.

\bibitem[{Feuerbacher} and {Caillard}(2004)]{itapdb:Feuerbacher2004}
{\sc {Feuerbacher},~M.,} and {\sc {Caillard},~D.},
\newblock 2004, {\it Acta Mater.}, {\bf 52}, 1297.

\bibitem[{Feuerbacher} et~al.(1997)]{itapdb:Feuerbacher1997b}
{\sc {Feuerbacher},~M., {Metzmacher},~C., {Wollgarten},~M., {Urban}~K., {Baufeld},~B.,
{Bartsch},~M.,} and {\sc {Messerschmidt},~U.},
\newblock 1997, {\it Mater.\ Sci.\ Eng.}, {\bf A233}, 103.

\bibitem[{Frenckel}, {Henley}, and {Siggia}(1986)]{itapdb:Frenkel1986a}
{\sc {Frenkel},~D.M., {Henley},~C.H.,} and {\sc {Siggia},~E.D.},
\newblock 1986, {\it Phys.\ Rev.}~B, {\bf 34}, 3649.

\bibitem[{Gratias}, {Katz}, and {Quiquandon}(1995)]{itapdb:Gratias1995}
{\sc {Gratias},~D., {Katz},~A.,} and {\sc {Quiquandon},~M.},
\newblock 1995, {\it J.~Phys.~Cond.~Matter}, {\bf 7}, 9101.

\bibitem[{Katz} and {Duneau}(1986)]{itapdb:Katz1986a}
{\sc {Katz},~A.} and {\sc {Duneau},~M.},
\newblock 1986, {\it J.~Phys.\ France}, {\bf 47}, 181.

\bibitem[{Katz} and {Gratias}(1994)]{itapdb:Katz1994d}
{\sc {Katz},~A.} and {\sc {Gratias},~D.},
\newblock 1994, {\it Lectures on quasicrystals}, edited by F.~{Hippert} and
D.~{Gratias} (Les Ulis: Les Editions de Physique), p. 187.

\bibitem[{Klein}(1997)]{itapdb:Klein1997c}
{\sc {Klein},~H.},
\newblock 1997, PhD Thesis, Institut National Polytechnique de Grenoble, France.

\bibitem[{Klein} et~al.(1996)]{itapdb:Klein1996}
{\sc {Klein},~H., {Audier},~M., {Boudard},~M., {de~Boissieu},~M., {Beraha},~L.,}
and {\sc {Duneau},~M.},
\newblock 1996, {\it Phil.\ Mag.}~A, {\bf 73}, 309.

\bibitem[{Klein}, {Durand}, and {Audier}(2000)]{itapdb:Klein2000a}
{\sc {Klein},~H., {Durand-Charre},~M.,} and {\sc {Audier},~M.},
\newblock 2000, {\it J.~All.\ Comp.}, {\bf 296}, 128.

\bibitem[{Klein} and {Feuerbacher}(2003)]{itapdb:Klein2003}
{\sc {Klein},~H.,} and {\sc {Feuerbacher},~M.},
\newblock 2003, {\it Phil.\ Mag.}, {\bf 83}, 4103.

\bibitem[{Klein} et~al.(1999)]{itapdb:Klein1999}
{\sc {Klein},~H., {Feuerbacher},~M., {Schall},~P.,} and {\sc {Urban},~K.},
\newblock 1999, {\it Phys.\ Rev.\ Lett.}, {\bf 82}, 3468.

\bibitem[{Koschella} et~al.(2002)]{itapdb:Koschella2002}
{\sc {Koschella},~U., {G{\"a}hler},~F., {Roth},~J.,} and {\sc {Trebin},~H.-R.},
\newblock 2002, {\it J.~All.\ Comp.}, {\bf 342}, 287.

\bibitem[{Letoublon} et~al.(2001)]{itapdb:Letoublon2001}
{\sc {L\'{e}toublon},~A., {de~Boissieu},~M., {Boudard},~M., {Mancini},~L.,
{Gastaldi},~J., {Hennion},~B., {Caudron},~R.,} and {\sc {Bellissent},~R.},
\newblock 2001, {\it Phil.\ Mag.\ Lett.}, {\bf 81}, 273.

\bibitem[{Mompiou}, {Caillard}, and {Feuerbacher}(2004)]{itapdb:Mompiou2004}
{\sc {Mompiou},~F., {Caillard},~D.,} and {\sc {Feuerbacher},~M.},
\newblock 2004, {\it Phil.\ Mag.}, {\bf 84}, 2777.

\bibitem[{Rosenfeld} et~al.(1995)]{itapdb:Rosenfeld1995}
{\sc {Rosenfeld},~R., {Feuerbacher},~M., {Baufeld},~B., {Bartsch},~M.,
{Wollgarten},~M., {Hanke},~G., {Beyss},~M., {Messerschmidt},~U.,} and {\sc
{Urban},~K.},
\newblock 1995, {\it Phil.\ Mag.\ Lett.}, {\bf 72}, 375.

\bibitem[{Socolar}, {Lubensky}, and {Steinhardt}(1986)]{itapdb:Socolar1986b}
{\sc {Socolar},~J.E.S., {Lubensky},~T.C.,} and {\sc {Steinhardt},~P.J.},
\newblock 1986, {\it Phys.\ Rev.}~B, {\bf 34}, 3345.

\bibitem[{Sun} and {Hiraga}(1996)]{itapdb:Sun1996a}
{\sc {Sun},~W.} and {{Hiraga},~K.},
\newblock 1996, {\it Phil.\ Mag.}~A, {\bf 73}, 951.

\bibitem[{Tsai} et~al.(1991)]{itapdb:Tsai1991a}
{\sc {Tsai},~A.P., {Yokoyama},~Y., {Inoue},~A.,} and {\sc {Masumoto},~T.},
\newblock 1991, {\it J.~Mater.\ Res.}, {\bf 6}, 2646.

\end{thebibliography}}

\end{document}